\documentclass[preprint]{aastex}

\def\simless{\mathbin{\lower 3pt\hbox
   {$\rlap{\raise 5pt\hbox{$\char'074$}}\mathchar"7218$}}} 
\def\simgreat{\mathbin{\lower 3pt\hbox
   {$\rlap{\raise 5pt\hbox{$\char'076$}}\mathchar"7218$}}} 

\def\be{\begin{equation}}
\def\ee{\end{equation}}

\def\thetaom{\mbox{\boldmath $\theta$}_{o,m} }
\def\thetaon{\mbox{\boldmath $\theta$}_{o,n} }
\def\vecu{\mbox{\boldmath $u$} }

\shorttitle{Polarized ISS of PKS 0405-385}
\shortauthors{Rickett, Kedziora-Chudczer \& Jauncey}

\received{2002 March 4}
\begin{document}

\title{Interstellar Scintillation of the Polarized Flux Density in Quasar, PKS 0405-385}
\author{Barney J. Rickett,}
\affil{Department of Electrical and Computer Engineering,
University of California, San Diego, CA 92093 and Australia Telescope National Facility}
\email{bjrickett@ucsd.edu}
\author{ Lucyna Kedziora-Chudczer}
\affil{Anglo-Australian Observatory and Australia Telescope National Facility}
\email{lkedzior@atnf.csiro.au}
\and
\author{David L. Jauncey}
\affil{Australia Telescope National Facility}
\email{djauncey@atnf.csiro.au}

\begin{abstract}
The remarkable rapid variations in radio flux density and polarization
of the quasar PKS 0405-385 observed in 1996 are 
subject to a correlation analysis, from which
characteristic time scales and amplitudes are derived. 
The variations are interpreted as interstellar scintillations.
The cm wavelength observations are in the weak scintillation
regime for which models for the various auto- and 
cross-correlations of the Stokes parameters are derived and fitted to
the observations. These are well modelled
by interstellar scintillation (ISS) of a $30 \times 22\mu$as source,
with about 180 degree rotation of the polarization 
angle along its long dimension. 
This success in explaining the remarkable intra-day variations (IDV)
in polarization
confirms that ISS gives rise to the IDV in this quasar.
However, the fit requires the scintillations to be occurring much
closer to the Earth than expected according to the standard
model for the ionized interstellar medium (IISM). Scattering at 
distances in the range 3-30 parsec are required to explain the observations.
The associated source model has a peak brightness temperature
near $2 \times 10^{13}$K, which is about twenty-five times smaller than
previously derived for this source.  This reduces the implied Doppler factor
in the relativistic jet, presumed responsible to 10-20, high but
just compatible with cm wavelength VLBI estimates for the Doppler factors in 
Active Galactic Nuclei (AGNs).
\end{abstract}


\section{Introduction}
\label{sec:intro}

Fluctuations on times of a day or less in the flux density 
of compact, flat spectrum  extragalactic 
sources at GHz frequencies were first 
discovered by Heeschen (1984). This, so called, 
{\em flickering} of radio sources was attributed 
to the flux density scintillation in the interstellar 
medium of our Galaxy (Heeschen \& Rickett 1987). 
The subsequent discoveries of much stronger and more rapid 
intraday variability (IDV) in a number of AGN (Witzel et al.\ 1986; 
Kedziora-Chudczer et al.\ 1997 [KCJ]; 
Dennett-Thorpe and de Bruyn 2000a [DTB];
Quirrenbach, et al.\ 2000;
Qian et al.\ 2000, 2001), 
ignited a debate as to their origin.
Many AGN radio sources were already known to show 
strong variability, though a thousand times slower, which was 
interpreted as intrinsic to the source. Therefore intrinsic 
processes were originally used to explain IDV as well.

However an intrinsic interpretation of strong 
IDV in extragalactic radio sources leads to 
a high brightness temperature, far in excess 
of the $T_{B}= 10^{12}$ K Compton limit 
for the synchrotron emission (Kellermann 
\& Pauliny-Toth 1969). This problem was 
temporarily solved by postulating strong 
beamed emission from a relativistic jet in IDV sources 
(Quirrenbach et al. 1992). 
Radiation from a relativistic shock in 
such a jet was subsequently postulated 
to reduce the bulk Lorentz factor needed to explain the
short time scales (Qian et al., 1991).
However the IDV of quasar, PKS 0405-385 (KCJ), with 50\% flux 
density changes within an hour, would require 
a Doppler factor, D$\sim10^{3}$, to explain 
the inferred variability brightness temperature 
$T_{B}^{var}\sim 10^{21}$K of a source as small 
as $10^{-10}$~arcsec.

The explanation of IDV in terms of 
relativistic beaming with a
Doppler factor higher than $10^{2}$ is 
difficult due to the very high energy 
requirements in the source (Begelman, Rees \& Sikora 1994). 
In addition such extreme relativistic beaming is 
not supported by the VLBI observations of 
superluminal sources (Vermeulen \& Cohen, 1994;
Kellermann, et al., 2000).
 
Rickett et al.\ (1995) [R95] analyzed the IDV of one particular 
well-studied quasar (B0917+624) and concluded that it 
was due to ISS in the ionized medium of our Galaxy. Narayan (1992)
also noted that the intrinsic interpretation of IDV
implies sources so small that they should 
also show strong effects of scintillation at radio frequencies.

Though the brightness temperature derived 
from ISS models depends on the variability 
time scale, it also depends critically on the distance ($L$)
to the scattering plasma and on the level 
of intensity modulation observed.
In the two extreme IDV sources (PKS 0405-385 and J1819+385)
the modulation index (rms/mean flux density) was found to peak 
at tens of percent at a frequency near 5 GHz (KCJ and DTB). 
From this one can derive the angular size of the scintillating component
as approximately the same as that 
of the first Fresnel zone at this
critical frequency, which divides the 
strong and weak scattering region. 
For PKS 0405-385 KCJ estimated the size of 
the scintillating component, 
5 $\mu $arcsec, corresponding to a 
brightness temperature $T_{b} > 5\times 10^{14}$ K, 
by assuming the distance to the scattering region to be 500 pc
(from the model for the interstellar plasma of Taylor and Cordes, 1993 [TC93]) 
and assuming a velocity of 50 km s$^{-1}$ relative to the line of sight. 
Though such a brightness is over six orders of magnitude less than
derived under an intrinsic interpretation, a Doppler factor of
$\sim 1000$ is still needed, since the brightness deduced by ISS
methods scales linearly with $D$.  Thus the problem remains of
how to explain such high $D$ values (see Marscher, 1998).

Conclusive evidence for the ISS interpretation of IDV comes from 
the observations of the time delay in the variability pattern observed 
in PKS 0405-385 between the ATCA and VLA (Jauncey et al., 2000)
and between Westerbork and the VLA 
for  J1819+385 (Dennett-Thorpe and de Bruyn, 2002). Further,
annual changes of the time scale of variability due to orbital 
motion of the Earth have been observed for sources J1819+385 
(Dennett-Thorpe 2000) and B0917+624 (Rickett et al.\ 2001; 
Jauncey and Macquart, 2001).  Neither of these phenomena can be 
explained as intrinsic variation.

One argument against the ISS interpretation of IDV is based on 
the misconception that scintillations cannot cause variability in 
the fractional polarized flux density, 
which is observed in IDV sources. Although a single isolated 
point source cannot cause rapid changes in the degree or angle
of linear polarization, IDV in the polarized flux 
density can be caused by two or 
more polarized components with misaligned position angles, 
at least one of which scintillates (Quirrenbach et al. 1989, R95).
Simonetti (1991) also proposed polarized substructure viewed
through a thin (0.04 AU) shock with a high plasma density
(500 cm$^{-3}$) as the cause of a single rapid rotation
in the 6-cm polarization angle from 0917+624 observed
during ongoing IDV-ISS.
   
In this paper we present the linear 
polarization observations for the quasar, PKS 0405-385 obtained during 
a period when the source exhibited its most rapid and strongest IDV 
(Section~\ref{sec:obs}). We describe 
the methods of statistical data analysis in Section~\ref{sec:analysis}.
In Section~\ref{sec:iss} we introduce the theory of weak
scintillation, which requires a detailed model for the
source brightness distribution and also for
the distance, velocity and density spectrum of the scattering plasma.
We compare the shape and time scale of the auto-correlation of
total intensity with theory, assuming a single Gaussian 
scintillating component and an extended non-scintillating component.
We conclude that the scattering plasma is highly anisotropic and given 
its velocity relative to the Earth, we obtain a new observational 
constraint on the effective source size and scattering distance.

In Section~\ref{sec:iss.stokes} we use ISS theory for a source 
with a polarized brightness that differs from that
of its total brightness to obtain expressions for 
the mutual cross-correlations between the Stokes parameters 
($I,Q,\& U$).  In Section~\ref{sec:fitting},
after fixing the scattering distance and axial ratio of the plasma,
we fit the theoretical correlations to the observations and obtain a 
polarization model of the source on the scale of 10 $\mu$as.
Though the modelling is not unique, its success 
confirms that ISS can indeed explain
the remarkable and complex pattern of variability in this source. 
We discuss the implications for the source and the IISM
in Section \ref{sec:disc}.

\section{Observations}  
\label{sec:obs}

Intraday variability of the quasar, PKS 0405-385 was 
discovered during the IDV Survey of 
the Australia Telescope Compact Array
(ATCA), as described by Kedziora-Chudczer et al., (1997, 2001). 
Follow-up monitoring of the source in June 1996 
revealed the unique nature of these variations, when 
rapid and strong, quasi-periodic 
fluctuations were found at the four ATCA frequencies: 
8.6, 4.8, 2.4 and 1.4 GHz (KCJ).
The data were sampled on a 24 min cycle:
the two higher frequencies were sampled
every 70 seconds for about 14 min, switching to the two 
lower frequencies for about 10 min.
Figures 1 and 2 of KCJ showed the resulting
time series for total flux density at all four frequencies
for nearly 12 hr on June 8, 9 and 10, 1996.
Data were also recorded for several hours on June 
7th and 11th.  Here we concentrate on results from the
three days with full 12 hr observations.

The variability of the total flux density with a typical 
time scale of an hour was strongly correlated
between the two highest frequencies.  
At 2.4 and 1.4 GHz fluctuations 
were slower and their amplitude decreased with 
decreasing frequency. KCJ interpreted these data
in terms of ISS in the weak (at 8.6 \& 4.8 GHz) 
and strong (at 2.4 \& 1.4 GHz) scattering regimes. They 
found that the observed spectrum of modulation index  
agreed with a theoretical prediction based 
on the TC93 model of interstellar medium,
assuming that the fraction of the flux 
density in the scintillating source component 
was 15\% at all frequencies.

\begin{figure}[hbt]
\includegraphics[height=15cm,angle=-90]{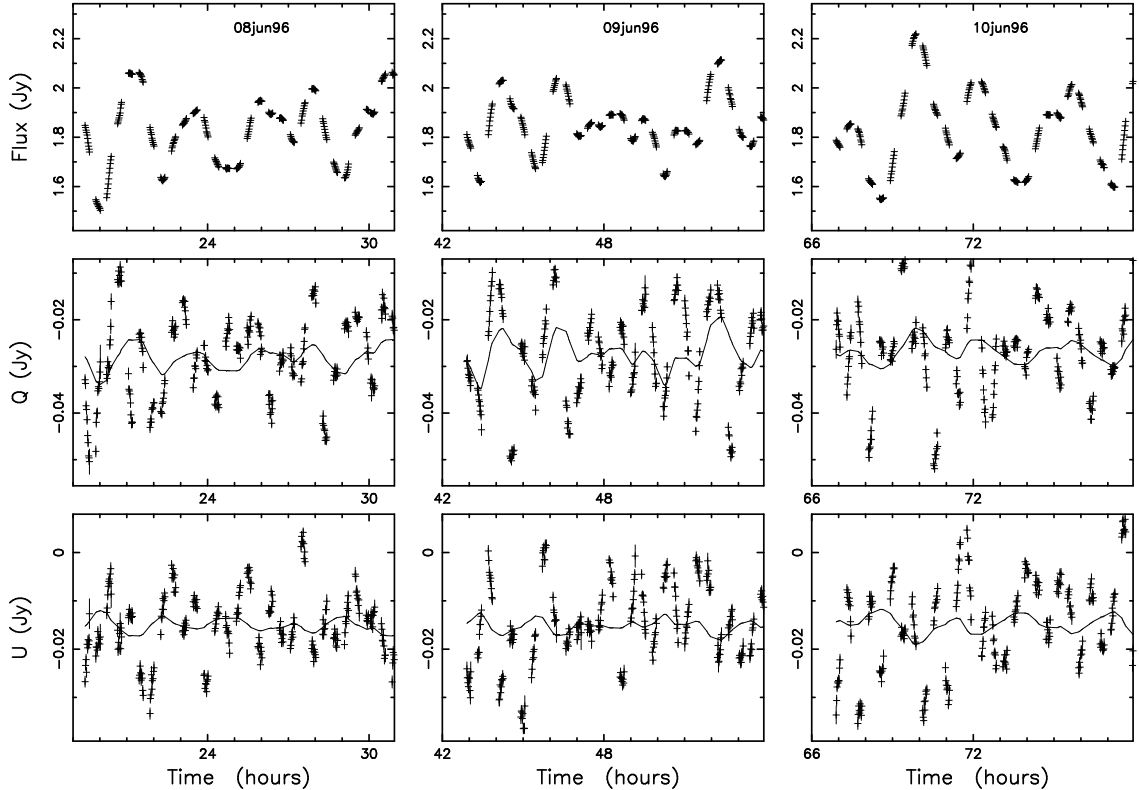}
\caption{Time series for $I,Q,U$ with $1\sigma$ errors at 8.6 GHz
for June 8-10, 1996. Times are relative to 0hr UT on June 7.  
The thin solid lines (best fits of the variations in $I$
to the variations in $Q$ and $U$) represent the
expected variability due a polarized point source or 
a linearly polarized extended source with uniform degree 
and angle of polarization.}
\label{fig:xIQU}
\end{figure}

The IDV in PKS 0405-385 continued for
about two months and then ceased and did not reappear
until November 1998.  The fastest time scale 
during this second episode was $\sim2$ times slower 
than the time scale observed in June 1996, 
and the highest amplitude of variability 
was half of that in the previous 
episode of IDV. The maximum amplitude
occurred in January 1999 
and the IDV ceased again in March 1999.  The 
transient behaviour of IDV in PKS 0405-385 
is presumed to be due to an expansion
of the scintillating component, which reduced 
the amplitude and increased the time scale of the ISS.

The long term monitoring of PKS 0405-385 
revealed that although the first period 
of IDV was followed by a brightening 
of the source at all frequencies 
(Kedziora-Chudczer et al. 2001), the 
second episode of IDV coincided with 
a gradual decrease in the average total 
flux density. As of October 2001, the source 
remains in a low state $\sim 1$ Jy at 4.8 GHz and is quiescent.
The goal of the present paper is to discuss
the observations from June 1996 and 
to present a detailed interpretation of the variations
including the very rapid IDV in linear polarization.

\begin{figure}[htb]
\includegraphics[height=15cm,angle=-90]{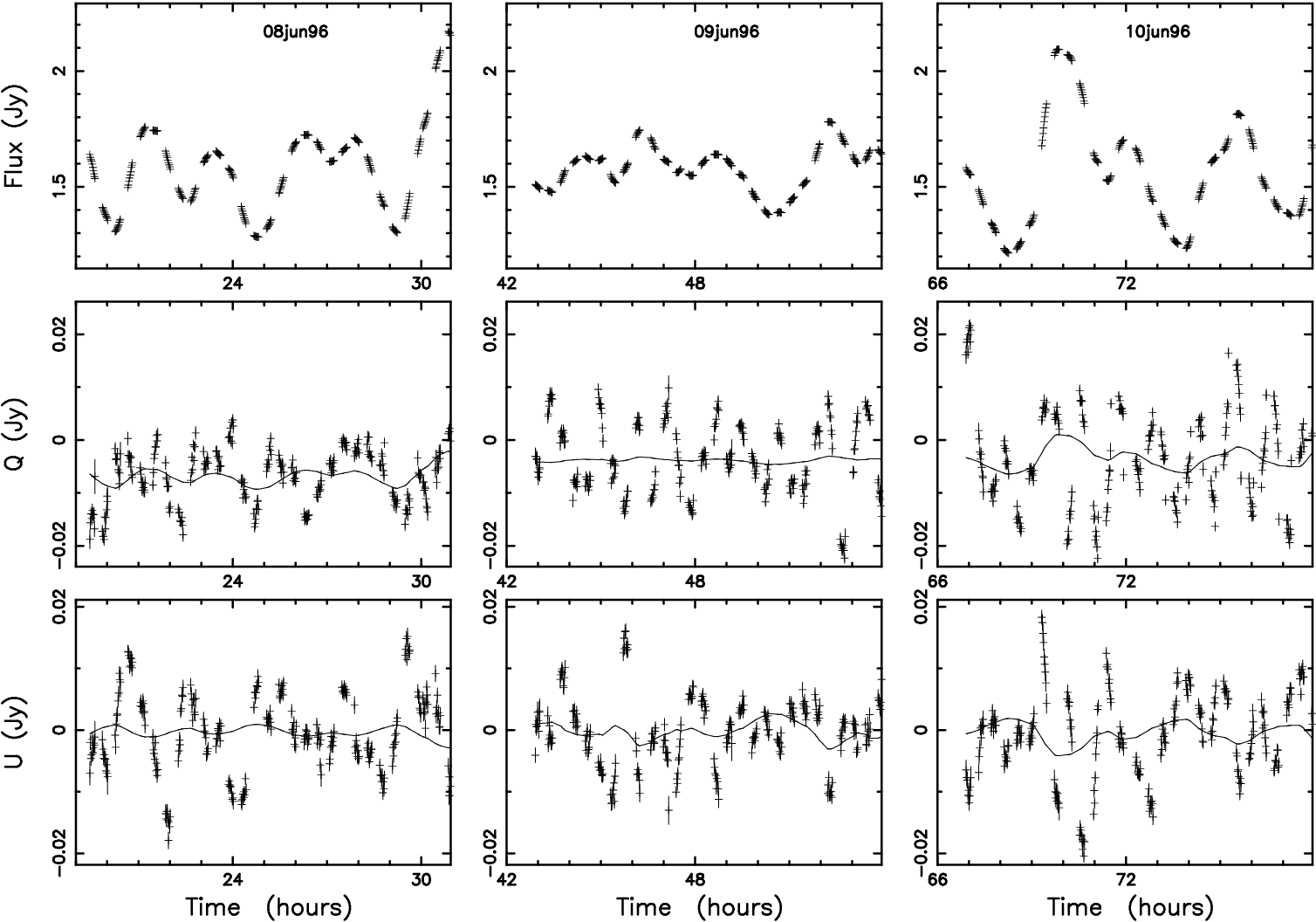}
\caption{Time series for $I,Q,U$ with $1\sigma$ errors at 4.8 GHz
for June 8-10, 1996, in the same format as in Figure \ref{fig:xIQU}.}
\label{fig:cIQU}
\end{figure}

\subsection{Polarization calibration}

The variations in linear polarization of PKS 0405-385 were measured with 
the ATCA at all four frequencies. 
The observations were performed with 10 sec 
integration time in continuum mode with 128 MHz bandwidth 
divided into 16 channels each from the X and Y polarizations. 
This configuration enables measurements of four polarization
products (XX,YY,XY and YX) for each frequency.
The three Stokes parameters $I,Q \&U$ were determined by calibration of the 
complex gains of the two polarization channels of the array. The instrumental 
leakages were corrected for by using a primary flux density calibrator, 
PKS 1934-638, which is known to be unpolarized (Komesaroff et al. 1984). 

The phase calibration is routinely done for each polarization channel 
separately. To combine information from polarization channels it is necessary 
to know the phase difference between channels. In the ATCA this difference 
is measured with a noise diode by injecting a signal into two polarization 
channels synchronously.
The polarization leakage terms are used to correct the complex gain solutions.  
The gains are applied to the visibilities to solve for Stokes $Q, U$ as a 
function of time as follows:
\begin{eqnarray}
 I & = & \frac{1}{2}(XX+YY), \nonumber \\
 Q & = & \frac{1}{2}[(XX-YY)\cos (2\chi_{\rm par} )-(XY+YX)\sin (2\chi_{\rm par} )], \\
 U & = & \frac{1}{2}[(XX-YY)\sin (2\chi_{\rm par})+(XY+YX)\cos (2\chi_{\rm par})], \nonumber
\label{eq:IQU}
\end{eqnarray}
where the linear feeds (X,Y) are corrected for instrumental leakages. 
The parallactic angle, $\chi_{\rm par}$ varies as a function of time.

\begin{figure}[htb]
\includegraphics[height=15cm,angle=-90]{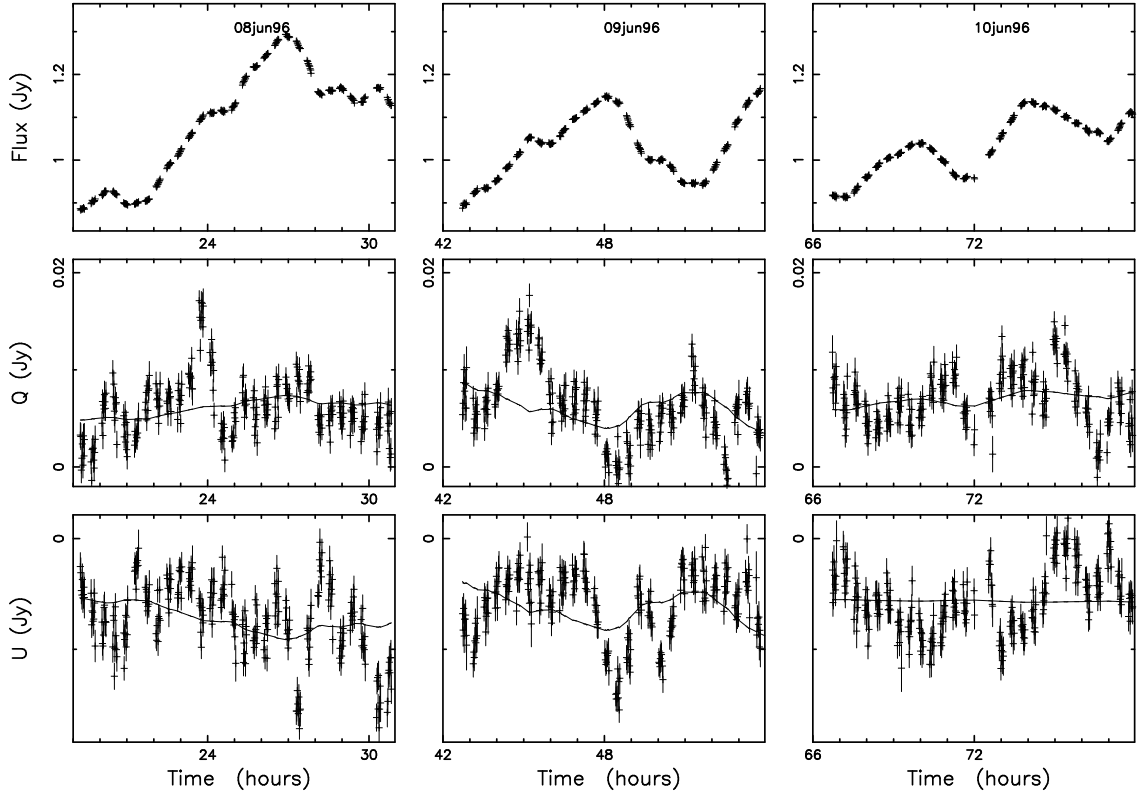}
\caption{Time series for $I,Q,U$ with $1\sigma$ errors at 2.4 GHz
for June 8-10, 1996, in the same format as in Figure \ref{fig:xIQU}.}
\label{fig:sIQU}
\end{figure}

\begin{figure}[htb]
\includegraphics[height=15cm,angle=-90]{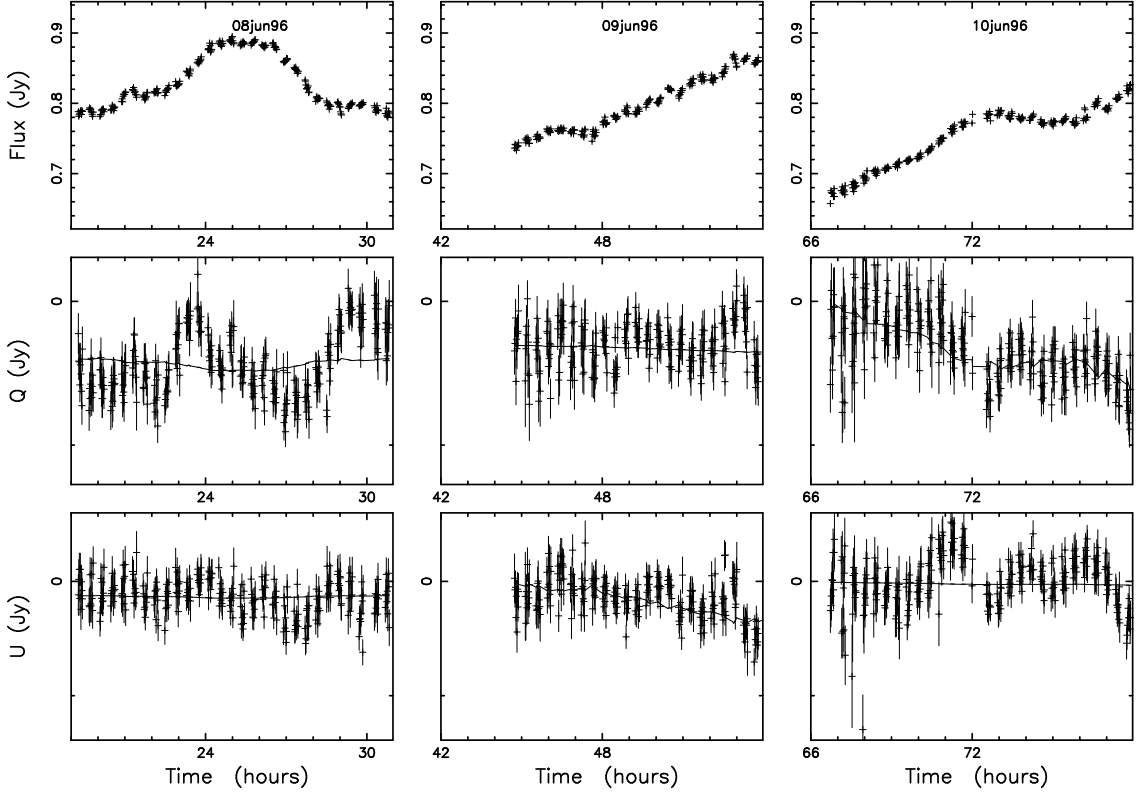}
\caption{Time series for $I,Q,U$ with $1\sigma$ errors at 1.3 GHz
for June 8-10, 1996, in the same format as in Figure \ref{fig:xIQU}.}
\label{fig:lIQU}
\end{figure}

The rms error of the polarized flux density measured over the 
observing session is a function of instrumental leakages and gain 
calibration. The instrumental polarization remains effectively constant
for the ATCA, 
therefore the main contribution to the observed rms comes from the
system gain stability. 
The 70 second estimates of $I,Q$ or $U$ were found by averaging the
10 second estimates from typically 8  central frequency channels
(8 MHz each) and up to 15 baselines.
The rms in these 10 sec estimates allowed an rms error to be estimated
for each 70 second sample. This error was less than 4 mJy at all 
frequencies, as determined for the large sample of the flat spectrum, 
compact sources (Kedziora-Chudczer et al. 2001). 
Figures \ref{fig:xIQU}, \ref{fig:cIQU}, \ref{fig:sIQU} and \ref{fig:lIQU} 
show three 12-hr observations of $I$, $Q$ and $U$ at 8.6, 4.8, 2.4 and 1.4 GHz,
respectively.  The errors are plotted as vertical bars
where they are larger than the height of the symbol.

\section{Data Analysis}
\label{sec:analysis}

\begin{figure}[tbh]
\plottwo{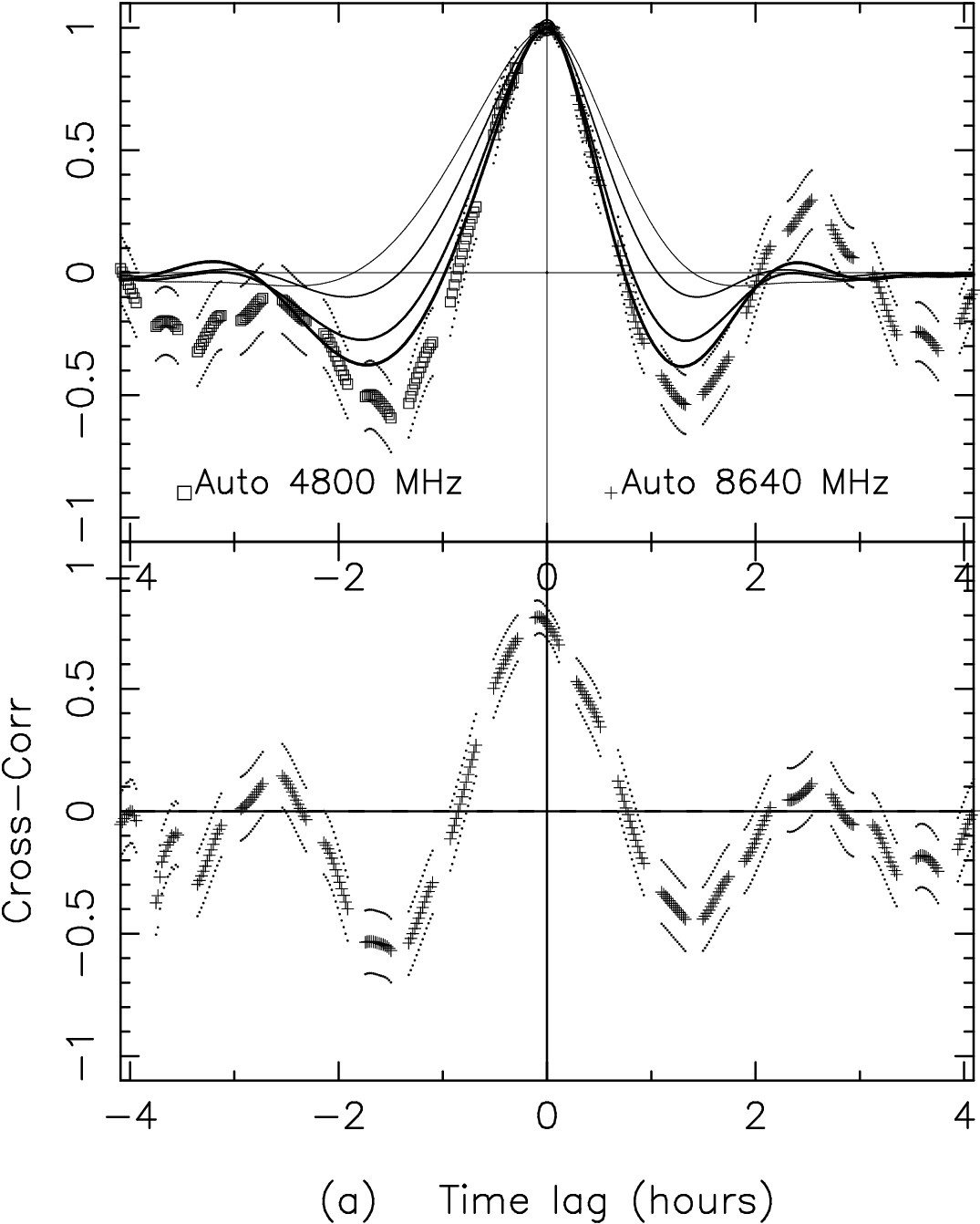}{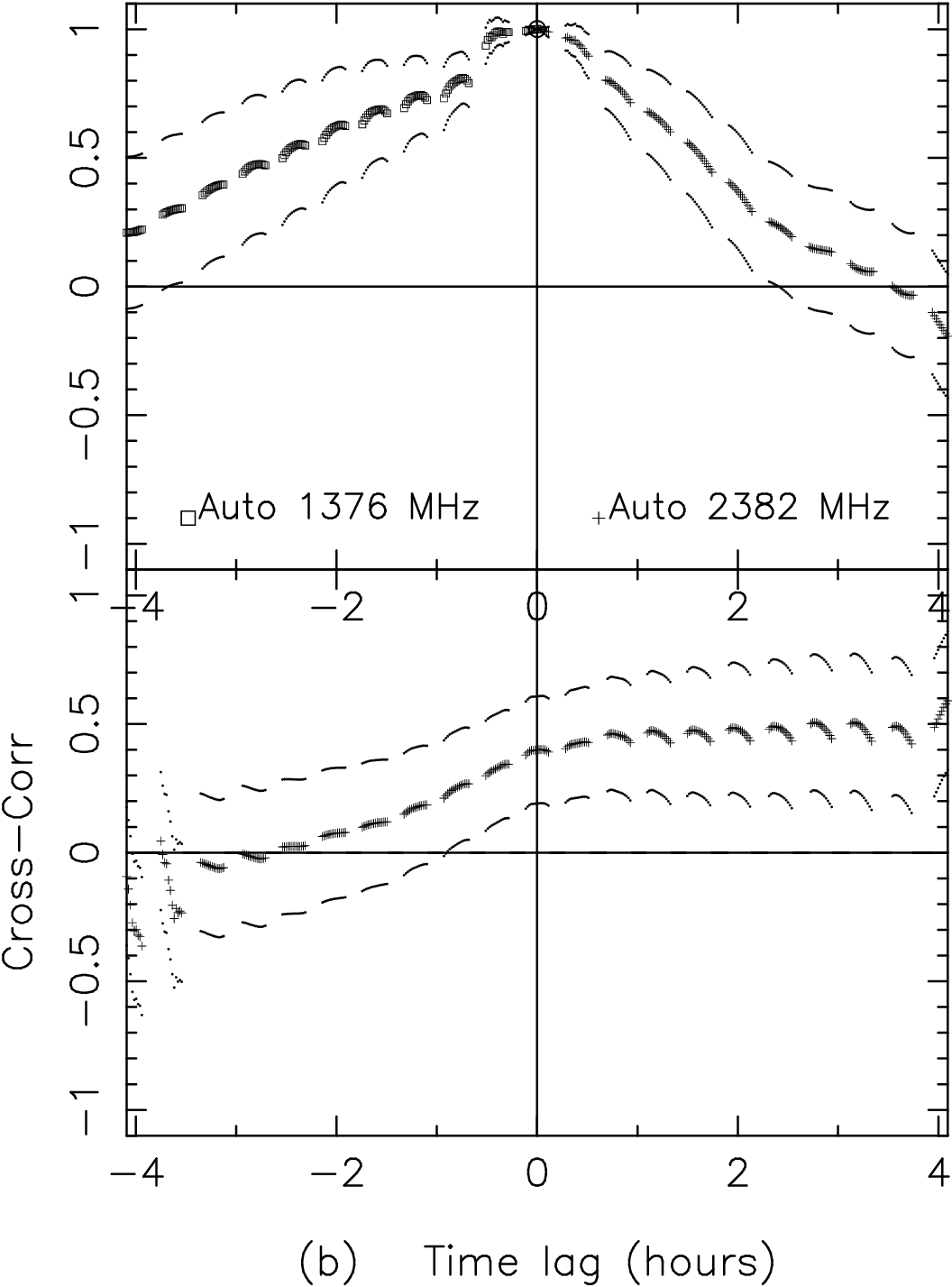}
\caption{Auto- and cross-correlations of total intensity for two pairs
of frequencies: (a) 8640 and 4800 MHz; (b) 2382 and 1376 MHz; 
the auto-correlation at the lower frequency 
of each pair is plotted against negative time lag. 
The $\pm 1\sigma$ errors are plotted as dots above and below.
In (a) theoretical acfs curves are overplotted for 
screen models by lines, whose thickness indicates the 
axial ratio from thickest to thinnest: 0.25, 0.5, 1.0 and 2.0,
as described in the text. For the cross correlations a positive 
lag indicates a variation observed at the higher frequency 
before the lower frequency.}
\label{fig:iccfs}
\end{figure}

Overall the variations in $I,Q$ and $U$ appear to be stochastic 
with no identifiable ``events'',
and so we characterize them by their various auto- and cross-correlations.
As noted by KCJ and, as can be seen in the figures,
the $I$-variations are clearly correlated between 8.6 and 4.8 GHz.
This is quantified by the auto- and cross-correlations plotted in
Figure \ref{fig:iccfs}a. At 2.4 and 1.4 GHz the $I$-variations are
more than a factor two slower but are not correlated
with the higher frequencies and only partially correlated with each
other as shown by their auto- and cross-correlations 
in Figure \ref{fig:iccfs}b.  

Since the samples were taken on a nearly periodic cycle with
gaps while the other frequency pair was sampled, there are gaps
in the time lags available for display. The method used was to
exclude time lags where the number of available products was 
less than one third of the number of samples.
The method for correcting the correlation functions 
for additive noise is described in Appendix A, where we also discuss
how we estimate the errors.

\begin{figure}[hbt]
\begin{tabular}{c}
\includegraphics[height=9cm,angle=-90]{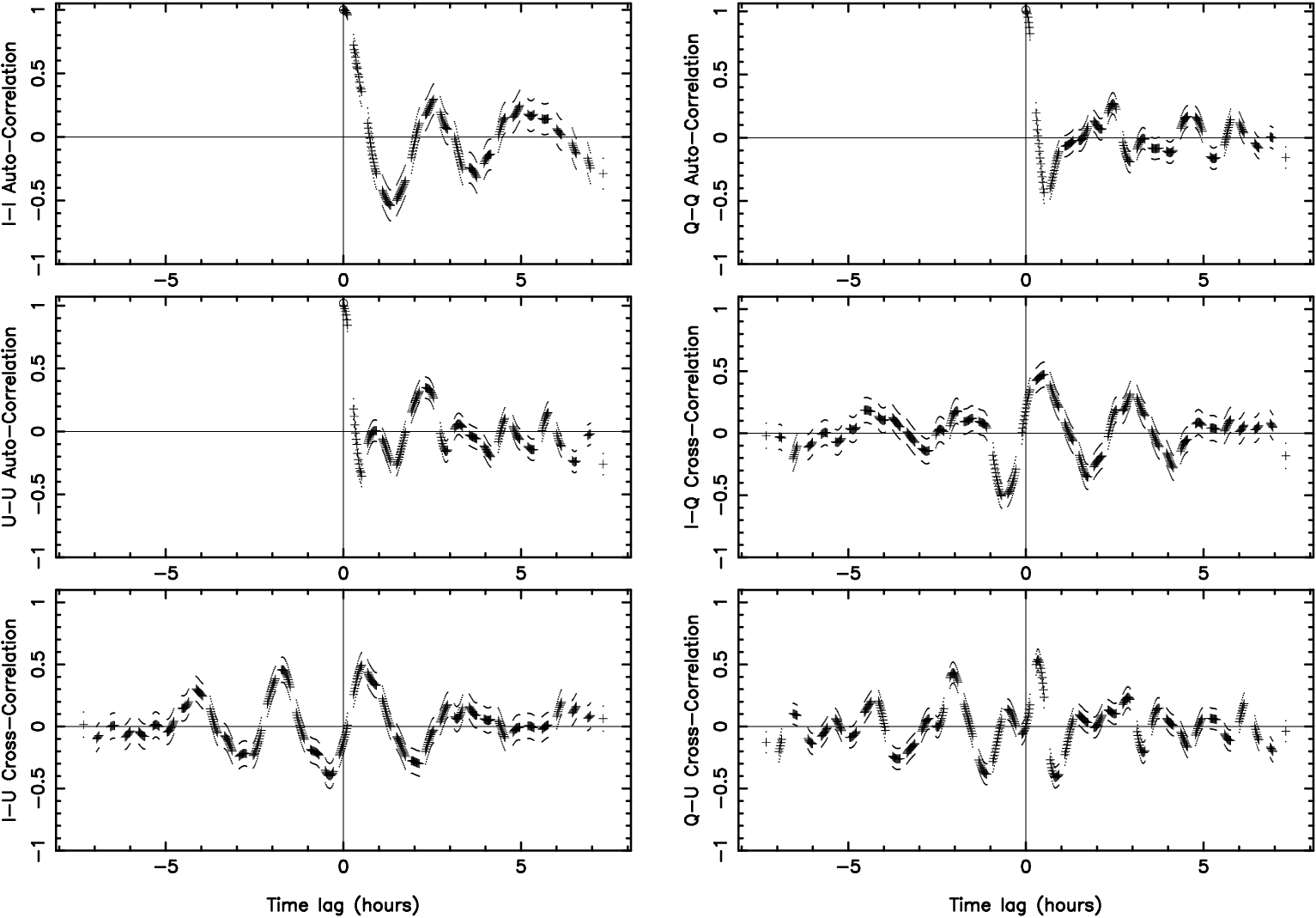} \\
\includegraphics[height=9cm,angle=-90]{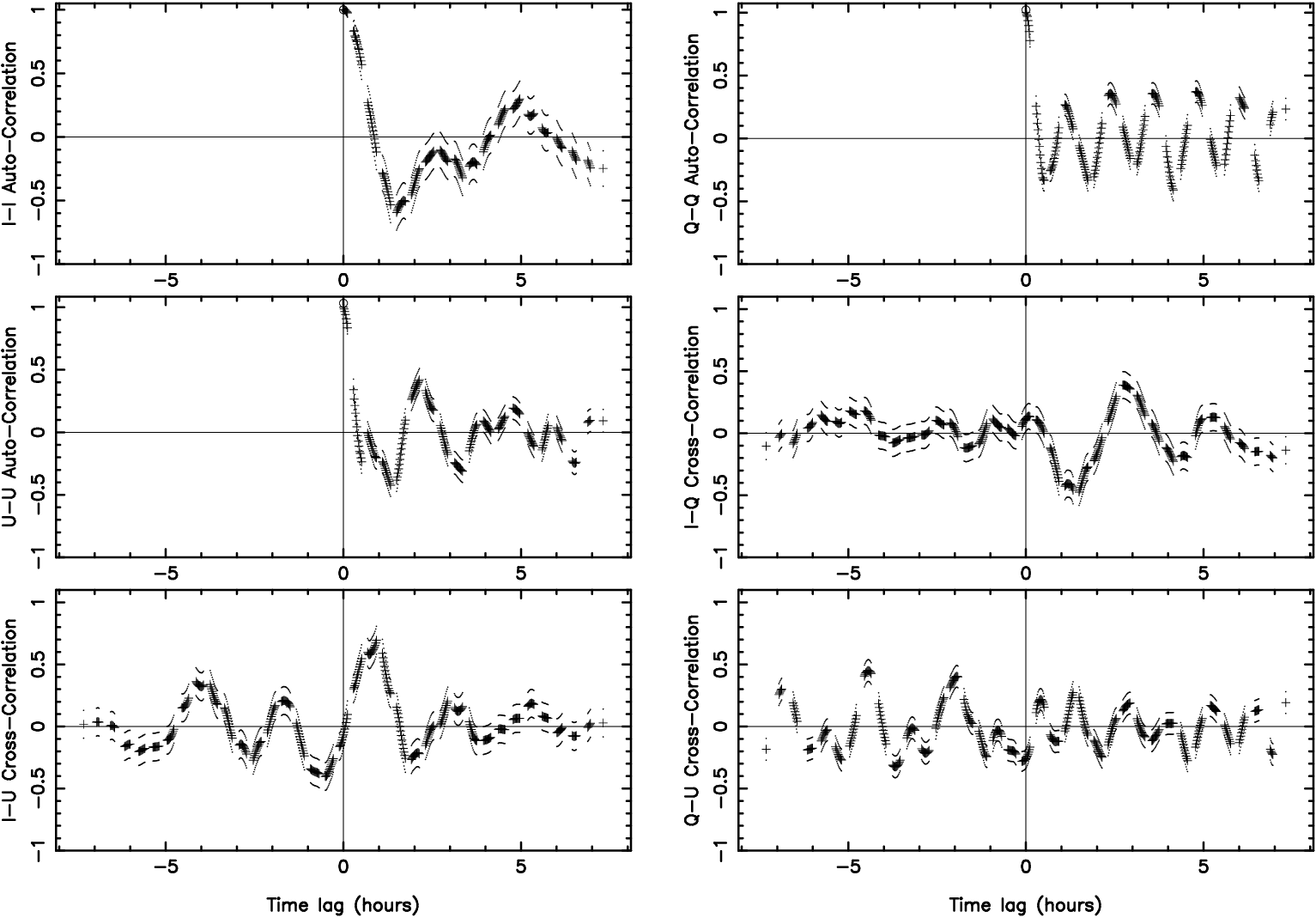}
\end{tabular}
\caption{Auto- and cross-correlations for $I,Q,U$
with $1\sigma$ errors indicated by dots above and below.  
Upper panels at 8.6 GHz, lower panels at 4.8 GHz.
Data were combined from June 8,9 \& 10 June 1996}
\label{fig:xc.IQU.cor}
\end{figure}

\begin{figure}[bht]
\begin{tabular}{c}
\includegraphics[height=9cm,angle=-90]{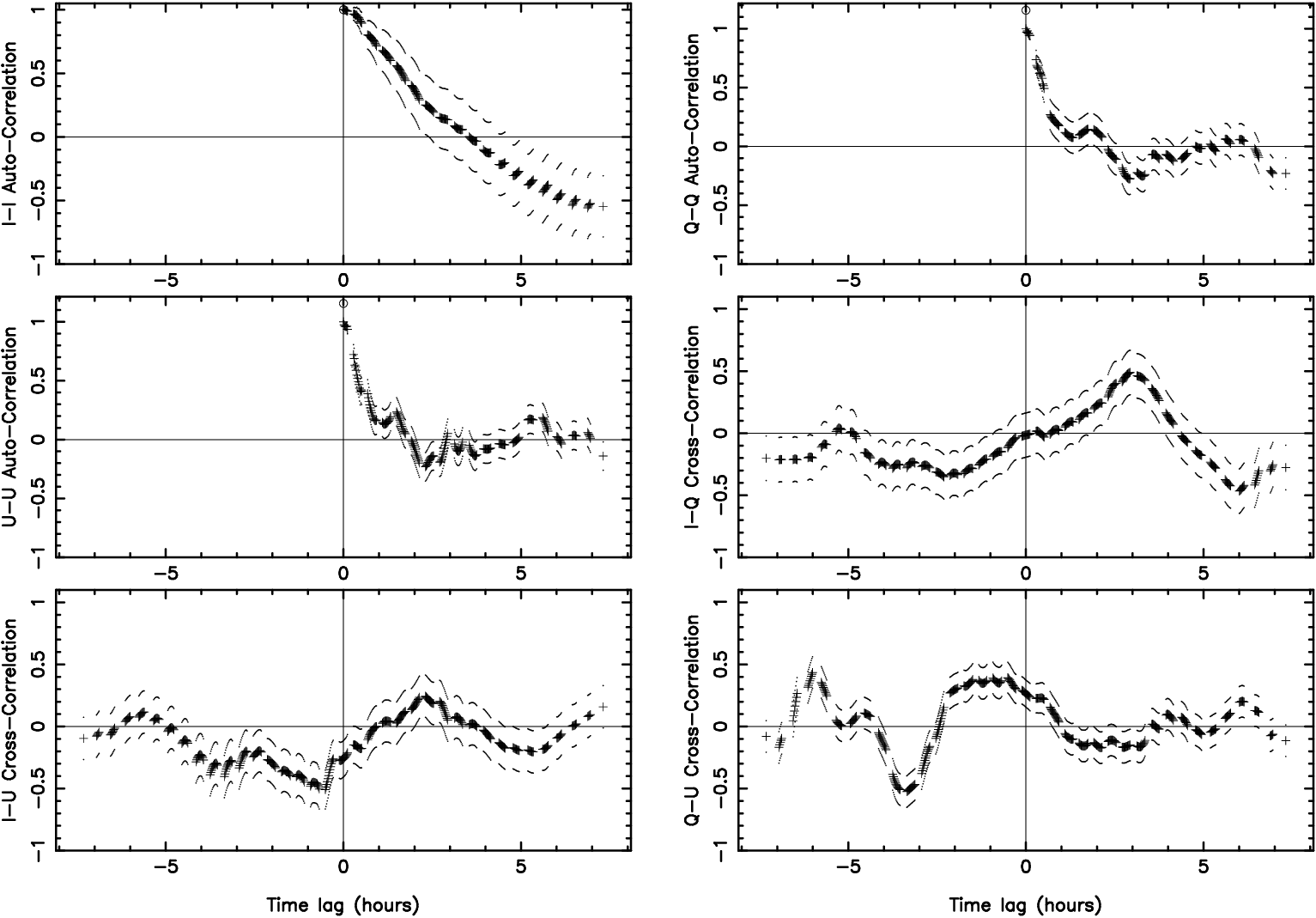} \\
\includegraphics[height=9cm,angle=-90]{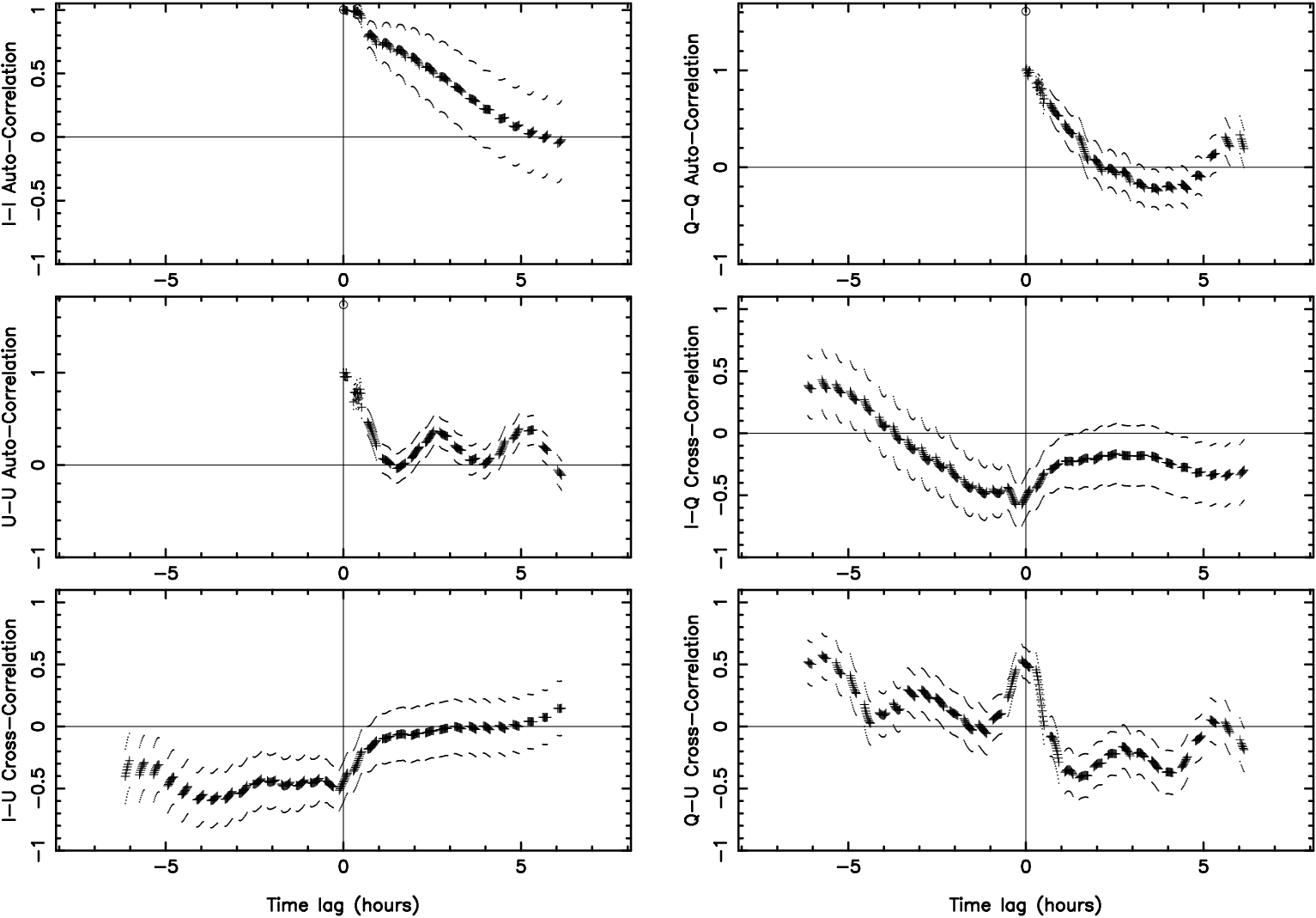}
\end{tabular}
\caption{Auto- and cross-correlations for $I,Q,U$, 
with $1\sigma$ errors indicated by dots above and below.  
Upper panels at 2.4 GHz, lower panels at 1.4 GHz.
Data were combined from June 8,9 \& 10 June 1996}
\label{fig:sl.IQU.cor}
\end{figure}

Figures \ref{fig:xIQU} to \ref{fig:lIQU}
show that the variations in $Q$ and $U$ are
clearly faster than in $I$ at each frequency,
with some signs of a relation between themselves and with $I$.
So we also computed the correlations between $I,Q$ at one 
frequency, defined in terms of 
the deviation $\Delta$ in each quantity from its mean:
\begin{eqnarray}
C_{II,t}(\tau) = < \Delta I(t) \Delta I(t+\tau) > \nonumber  \\
C_{QQ,t}(\tau) = < \Delta Q(t) \Delta Q(t+\tau) > \nonumber \\
C_{UU,t}(\tau) = < \Delta U(t) \Delta U(t+\tau) > \\
C_{IQ,t}(\tau) = < \Delta I(t) \Delta Q(t+\tau) > \nonumber \\ 
C_{IU,t}(\tau) = < \Delta I(t) \Delta U(t+\tau) > \nonumber \\ 
C_{QU,t}(\tau) = < \Delta Q(t) \Delta U(t+\tau) > \nonumber
\label{eq:Ctdefs}
\end{eqnarray}
In Figures \ref{fig:xc.IQU.cor} and \ref{fig:sl.IQU.cor} we show
all of these correlations for the four frequencies; 
the $\pm 1\sigma$ errors in the
individual points in these correlation functions are shown 
by dots above and below each symbol. 
As discussed in Appendix A, the errors are dominated by the \it estimation error \rm
with only small contributions from system noise. Thus
they are not independent from one time-lag to the next and are
correlated over a typical time scale for each data set.  This can be seen clearly
from the substantial ripples that remain at large time
lags (i.e.\ 3--8 hr).

We estimate the variances and characteristic time scales from each 
auto-correlation function, as tabulated with 
their errors in Table \ref{tab:rms.tau}.
We choose to define 
the characteristic time scale for each Stokes parameter
($\tau_{I}$, $\tau_{Q}$ \& $\tau_{U}$) as the HWHM
of its auto-correlation function, as used in R95.
There are several other alternative definitions for a
characteristic time scale.
KCJ used the average time interval between peaks in the
observed flux density time series, reporting 2 hrs
as the time scale at 4.8 GHz, compared with 
0.55 hrs from our definition. As we shall see later,
the time for the first minimum in the auto-correlation
function is another important time scale
(which is about three times $\tau_{I}$ at 4.8 GHz).
Whereas the definition is simply a matter
of convention, it must be applied consistently to
both observation and theory, which we do in 
section \ref{sec:iss}.

\clearpage


We have also computed the cross-correlations 
in $I$ between neighbouring frequency 
bands and tabulated the correlation coefficients at 
zero-lag in Table \ref{tab:icor.freq}. 
Since the sampling at the low and high frequencies were interleaved
the zero-lag correlation between 4.8 and 2.4 GHz was found
by interpolation.


\begin{table}[htb]
\begin{tabular}{lcccccccc}
\tableline
\renewcommand{\arraystretch}{0.6}
Freq.& $<S>$& $S_{\rm min}$ & $I_{rms}$ & $Q_{rms}$ & $U_{rms}$ & $\tau_I$ & $\tau_Q$ & $\tau_U$ \\
GHz &  Jy & Jy & mJy & mJy & mJy & hr & hr & hr \\
\tableline
8.64 & 1.85 & 1.44 &  138 $\pm$ 8 &  9.5 $\pm$ 0.4 & 8.0 $\pm$ 0.3  
&  0.41 $\pm 0.06 $ & 0.20 $\pm$ 0.02  &  0.20 $\pm$  0.02 \\
4.80 & 1.58 & 1.21 &  185 $\pm$ 13 & 7.4 $\pm$ 0.3 & 6.3 $\pm$ 0.3  
&  0.55 $\pm 0.07 $ & 0.20 $\pm$ 0.02  &  0.22 $\pm$  0.02     \\
2.38 & 1.05 & 0.88 &  98 $\pm$ 12 &  3.1 $\pm$ 0.2 & 3.1 $\pm$ 0.2  
&  1.6 $\pm 0.6 $ & 0.5 $\pm$ 0.1  &  0.4 $\pm$  0.1     \\
1.38 & 0.80 & 0.66 &  51 $\pm$ 8 &  1.8 $\pm$ 0.2 & 1.7 $\pm$ 0.1  
&  2.6 $\pm 1.3 $ & 1.0 $\pm$ 0.4  &  0.7 $\pm$  1.0     \\
\tableline
\end{tabular}
\caption{ Estimates of the mean and minimum flux densities and the 
rms variation in Linear Stokes parameters 
and the associated time scales.}
\label{tab:rms.tau}
\end{table}


\begin{table}[htb]
\begin{tabular}{ccc}
\tableline
Frequency & Frequency & $\rho$ \\
GHz &  GHz & \\
\tableline
8.64 &  4.80 &  0.77 $\pm$ 0.06   \\
4.80 &  2.38 &  0.2 $\pm$ 0.2    \\
2.38 &  1.38 &  0.4 $\pm$ 0.2    \\
\tableline
\end{tabular}
\caption{ Correlation coefficients for the intensity between
pairs of frequencies}
\label{tab:icor.freq}
\end{table}

The Stokes' correlations were computed by combining the three 12-hr tracking
observations on June 8, 9 and 10, 1966.  We also examined these
correlations from each individual day.  The daily
8.6 GHz cross-correlations 
of $IQ$, $IU$ and $QU$ were of the same general form
as in the three day average (Figure \ref{fig:xc.IQU.cor}a).
In particular the time offsets between the $IQ$ and $IU$
correlations were evident in each day, as was the 
narrow peak in $QU$ at a time lag of about 20 min. 
However, in contrast the Stokes' cross-correlations at 
4.8 GHz were not consistent over the three days, with the exception of
the persistence of a peak in $IU$ at about 1 hr which is
visible in the three-day average.  In comparing the 
rms amplitudes in Table \ref{tab:rms.tau}, we see that 
the 4.8 GHz $I_{rms}$ is 30\% larger than at 8.6 GHz, but
$Q_{rms}$ and $U_{rms}$ are lower by 25\%.  Thus fractionally
the polarization IDV at 4.8 GHz is well below that at 8.6 GHz.
This suggests that there may be a greater depolarization
of the linear polarized structure at 4.8 GHz
than at 8.6 GHz.

\begin{figure}[tbh]
\includegraphics[height=9cm,angle=-90]{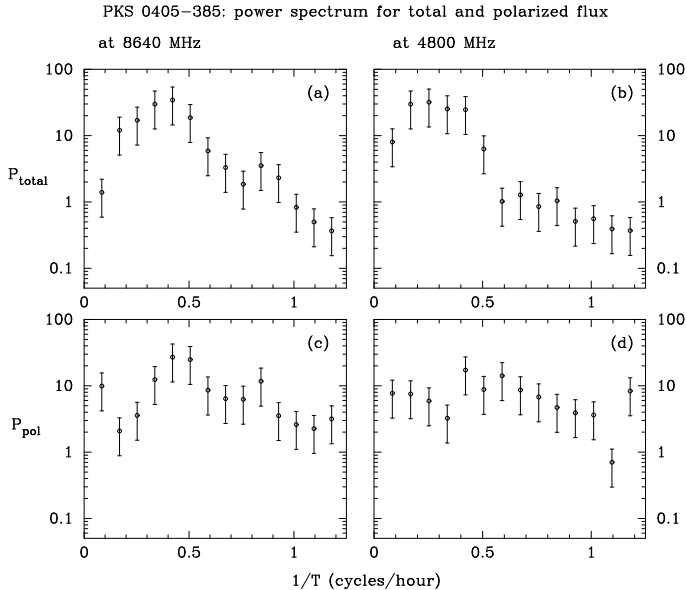} 
\caption{Temporal power spectrum for PKS 0405-385 
averaged from data on June 8, 9 and 10. (a) and (c) are at 8.6 GHz;
(b) and (d) are at 4.8 GHz. (a) and (b) show power spectrum 
for $I$; (c) and (d) are for $P$. }
\label{fig:powerspec}
\end{figure}

In Figure \ref{fig:powerspec} we 
show the temporal power density spectrum (PDS) for
the total and polarized flux density at 4.8 and 8.6 GHz. The spectra were
computed using the Lomb (1976) method\footnote{We used the algorithm from the
Numerical Recipes Fortran library.} for the
analysis of unevenly sampled data. The PDS is the average 
of the spectra from June 8, 9 and 10. Since our data sampling 
was for 14 min out of every 24 min, there is 
an effective Nyquist frequency at 1.25 cycles per hr (cph).  
The total flux density spectra show clear 
peaks around 0.4 cph, corresponding to a period of 2.5 hr, and fall 
steeply out to the Nyquist frequency. Thus the spectra are
only slightly aliased. However, the variations in the polarized 
flux density are faster, and its
PDS extends to higher frequencies. As a consequence the spectra in 
Figure \ref{fig:powerspec}c and d are flatter and so are
more heavily aliased. The interpretation of the observed PDS 
features, in terms of anisotropic
scattering, is given in section \ref{sec:aniso}.

\section{Interstellar Scintillation}
\label{sec:iss}

The variations in total and polarized flux density 
have a stochastic appearance consistent with ISS
and so we model the variations as purely due to ISS,
with no intrinsic variation. We estimate correlation functions 
from the observations and compare them with computations
from ISS theory.  

A marked difference exists between weak and strong ISS 
(see Rickett 1990 and Narayan 1992 for reviews).  
Most information on the phenomena has been obtained
from pulsar measurements below about 2 GHz and from lines
of sight that are so heavily scattered that the angular broadening
can be observed with Earth-based VLBI.  These are observed in 
the strong scintillation regime, in which there are two distinct
time scales of variation in flux density - a diffractive time scale
$\tau_d$ in the range minutes to an hour characterized by 100\%
modulations (with narrow band properties)
and a refractive time scale $\tau_r$ in the range of hours 
to days over a broad-band with a smaller amplitude. The
two time scales are related to characteristic spatial scales
$s_d= V\tau_d$ and $s_r = V \tau_r$ via the relative velocity 
$V$ of the observer with respect to the diffraction pattern.
These scales are in turn related to the Fresnel scale $r_F$
\be
s_d  s_r = r_F^2 = L/k ,
\label{eq:scales}
\ee
where $k$ is the radio wavenumber and $L$ characterizes the distance 
to the scattering medium -- as a screen distance or as the path length
(to 1/e) for an extended scattering medium.  

As the observing frequency is raised on a given line of sight these two scales 
converge, and, at a transition frequency $f_w$ they become equal.  At still higher
frequencies the scintillations have a single time scale $\tau_I$
\be 
\tau_I = r_F/V ,
\ee
and are referred to as ``weak'', though the amplitude is approximately 100\%
at $f_w$ and decreases as an inverse power of frequency.
From the TC93 model the expected transition frequency is about 5 GHz 
for the line of sight to PKS 0405-385 
at -48 degrees Galactic latitude.
Note that this value was obtained from Walker (2001),
which corrected his 1998 calculation, adjusted for
our definition that $f_w$ is the frequency at which the IISM
causes one radian rms across the Fresnel scale.
Thus our two higher frequencies should be described by weak 
ISS and the two lower
frequencies by strong ISS. Although the standard model for
the distribution of the scattering plasma (TC93) 
is only approximate and there is substantial uncertainty
in $f_w$ on any particular line of sight,
the high correlation observed between 5 and 8 GHz confirms 
this division.

The influence of the source diameter on the visibility of the scintillations
is crucial in our interpretation, as it was in the work of 
R95 on their ISS interpretation of the
IDV from quasar 0917+624.  The basic point is as simple as
``stars twinkle, planets don't'' -- i.e.\ the
absence of optical atmospheric twinkling from planets whose
angular size is substantially greater than that of stars.
In the same way most radio sources subtend too large an
angular diameter to exhibit the effects of ISS, while the
angular size of pulsars is so small that they exhibit 
the full range of ISS phenomena.  Though the compact cores of 
active galactic nuclei (AGN) are larger in angular size than pulsars,
many are small enough to show some level of ISS.
In particular, the phenomenon of low frequency variability
is now thought to be due to refractive ISS, rather than intrinsic
to the source (Rickett, 1986). Low amplitude flickering
over a few days at 3.3 GHz 
is also thought to be ISS (Heeschen and Rickett, 1987).

The influence of source structure on scintillation can be described
simply when the scattering occurs in a localized region modelled
as a phase screen at a distance $L$ from the observer. 
If the emission is spatially incoherent, the scintillation 
intensity pattern from an extended source is the pattern from a point source
convolved by the source brightness distribution
scaled by the lever-arm $L$.
Accordingly, scintillations will be suppressed when the source size 
$\theta_{\rm so}$ is larger than the angular size of the relevant
scale in the diffraction pattern subtended at the observer.
Thus in strong ISS diffractive scintillations are suppressed if 
$\theta_{\rm so} \simgreat \theta_d$, where $\theta_d = s_d/L$.
Similarly refractive scintillations are suppressed if
$\theta_{\rm so} \simgreat \theta_r = s_r/L = 1/(k s_d)$.
In weak ISS the scintillations are 
suppressed if $\theta_{\rm so} \simgreat \theta_w = r_F/L$.
Thus depending on the observing frequency and
the effective diameters of the various components that make up a radio
source, it may exhibit ISS in any of the three regimes
(e.g.\ figure 1 of Rickett, 2001).
For a scattering plasma distributed along the line of
sight, similar conditions govern the observation of the three regimes;
however, the corresponding inequalities depend on the form 
of the scattering profile.

We know from radio imaging studies that the brightness
distributions of AGNs typically contain structures on
scales ranging from an arcsecond to the resolution limit of VLBI
(smaller than 1 milli-arcsecond - mas).  
%
The observed  parameters of typical IDV include rms amplitudes of 
only a few percent and time scales of hours to days (e.g.Quirrenbach et
al. 1992). These are consistent with ISS in which the
smallest source size is large enough to partially quench the 
weak and refractive scintillations and fully quench diffractive
scintillation. This agrees with the low upper limits on diffractive 
scintillation obtained by Condon and Backer (1975) and, in particular
for PKS 0405-385 by Kedziora-Chudczer, Jauncey and Macquart (2002). 
Consequently, we now assume the diffractive regime to be
entirely quenched, allowing us to use the 
simplified theory of Coles et al.\ (1987).
%

The overall goal of our analysis is to compare the 
IDV observed in PKS 0405-385 with quantitative predictions of ISS.
We first examine the range of possible ISS models for the 
IDV in total intensity and
postpone analysis of the polarization to
sections \ref{sec:iss.stokes} and \ref{sec:fitting} 


\subsection{An ISS model for IDV}
\label{sec:iss.idv}

Consider, first, a characterization of the IDV by its rms
amplitude in total flux density ($S_{\rm rms} = I_{\rm rms}$) 
and its characteristic time scale  ($\tau_I$) at each frequency.  
KCJ observed a clear maximum in the modulation index 
($S_{\rm rms}$/$<S>$) of 0.13 at 4.8 GHz, from which they
concluded that 4.8 GHz is close to the transition
frequency ($f_w$) which divides weak from strong scintillation.
Simulations have shown that weak ISS theory can be used
to a good accuracy, even near the transition frequency
(see the discussion in Rickett et al., 2000). Thus as
in KCJ we use weak ISS theory at both 4.8 and 8.6 GHz.

KCJ also made estimates of the source diameter and brightness
temperature, by assuming that the distance to the scattering plasma
is about 500 pc, based on the TC93 model.
In what follows we remove this assumption and explore what range of
scattering distance is compatible with the total intensity 
observations; in turn this provides 
an allowable range for the source diameter and brightness temperature.

Since the source may well have substantial flux density in components that
are too large to scintillate, but remain unresolved in VLBI,
the average flux density $S_c$ of the scintillating component is not known.
Since it can be no more than the total average flux density
$S_{\rm T}$ and cannot be negative, we find 
$S_{\rm T} \ge S_c \ge S_{\rm T}-S_{\rm min}$,
where $S_{\rm min}$ is the lowest flux density observed.
Consequently we can constrain the actual modulation index
of the compact core $m_c = S_{\rm rms}/S_c$ by:
\be
S_{\rm rms}/S_{\rm T} = m_{\rm app} \le m_c \le S_{\rm rms}/(S_{\rm T}-S_{\rm min})  \; .
\ee
Considering the 4.8 GHz observations from June 8-10 1996, we obtain:
$0.13 \le m_c \le 0.52$ for the scintillation index 
and constrain the time scale $0.45 \le \tau_I \le 0.65$ hrs.
At 8.6 GHz the limits are $0.08 \le m_c \le 0.32$ and 
$0.31 \le \tau_I \le 0.51$ hrs (see Table \ref{tab:rms.tau}).
We now investigate how these parameters of the ISS can be used to constrain the
source diameter and scattering distance.

Here we model the source as a single Gaussian component with compact flux 
density $S_c$
and a diameter (FWHM) $\theta_{\rm mas}$ in milli-arcsec
(in contrast to HWHM definition for time scale).
We model the density spectrum of the scattering plasma
by the Kolmogorov form (see Armstrong et al., 1995)
with anisotropy included:
\be
P_{ne}(\kappa_x,\kappa_y,\kappa_z) = C_N^2(z) (R \kappa_x^2 + \kappa_y^2 /R + \kappa_z^2 /R_z)^{-(\alpha+2)/2} ,
\label{eq:Pne}
\ee
where $\alpha = 5/3$ makes it a Kolmogorov spectrum (at wavenumber
$\kappa_x,\kappa_y,\kappa_z$),  and
the axial ratio is $R$ when projected onto the $x,y$ plane;
note that ISS is not sensitive to any longitudinal structure
represented by $\kappa_z$ and $R_z$ (see Backer and Chandran, 2002 for a
related discussion). 
We use the theory of weak ISS,
as given by Coles et al.\ (1987), excluding the refractive
cut-off in the spectrum and generalized to anisotropic scattering.

The profile of scattering strength 
is assumed to be either a thin screen at distance $L$
\be
C_N^2(z) = C_{N0}^2    \;\;\;\;\;\; L-\delta L/2 \le z \le L+\delta L/2
\ee 
or a Gaussian function of distance 
$z$ from the Earth with scale length $L_g$:
\be
C_N^2(z) = C_{N0}^2  \exp(-z^2/L_g^2) \; .
\label{eq:cnsq}
\ee  
The velocity of the scattering medium is another critical parameter, 
which is not well known.  We adopt a value of 36 km s$^{-1}$, which
is the velocity of the Earth relative to the local 
standard of rest (LSR) at the time of the June 1996 observations.
We note that this is consistent with the measurement
a time delay in the ISS between the VLA and ATCA, which
gave an upper limit to the velocity of 75 $\pm15$ km s$^{-1}$
(Jauncey et al., 2000).  Velocities close to that of the LSR
are also supported by the recognition of annual modulation
in the timescale of IDV in two sources
(Rickett et al., 2001, Jauncey \& Macquart, 2001 and Bignall, 2002).
A velocity of about 35 km s$^{-1}$ relative to the LSR was obtained
for the scintillations of quasar J1819+385
both from and intercontinental time delay  measurement
and from the annual modulation in that source
(Dennett-Thorpe and de Bruyn, 2002a,b).

\subsection{Anisotropic Scattering}
\label{sec:aniso}

With the foregoing model assumptions, we computed 
auto-correlation functions (acfs) at 8.6 GHz and 4.8 GHz for
particular screen models and compared them with the observations. 
We initially assumed isotropic scattering ($R=1$) and chose the screen distance
to match the width of the observed acf.
With a small enough source diameter we expect $\tau_I \sim r_F/V$,
which implies a screen distance $L \sim 16$ pc. 
However, the comparison showed that the observed acfs have a much 
deeper minimum than the theory. 
In Figure \ref{fig:iccfs}a the first minimum at time lag of
1.3-1.5 hrs is near -0.5, which is more than 4$\sigma$ below
the isotropic model. Consequently, we also explored anisotropic models,
since they can explain this phenomenon which we term a negative ``overshoot''.

In the upper left panel of Figure \ref{fig:iccfs}
the observed acfs at 4.8 and 8.6 GHz are overplotted with
computed curves for axial ratios: 2.0, 1.0, 0.5 and 0.25. 
The axial ratio 1.0 corresponds to isotropic scattering,
for the ratio 2.0  the orientation of the velocity 
vector is parallel to the long axis
of the plasma irregularities, 
while for 0.5 and 0.25 it is parallel to the short axis.
The solid line with the smallest axial ratio 
(greatest anisotropy) gives the best match
to the depth of the negative overshoot and also 
the narrowest width.  
Consequently, in these plotted curves we increased the
screen distance to 25 pc, to match the observed width
for an axial ratio of 0.25. 
Changing the ellipse orientation from parallel 
to perpendicular changes the shape smoothly from the
dashed to the dotted curves. Note that there is a quite wide 
range ($\pm45$ degrees) in orientation angle 
over which a substantial negative overshoot is obtained,
as can be seen in Figure \ref{fig:ptspread} which
shows the point source correlation function in two dimensions.
In the calculation the source diameter was set to be small enough that
it does significantly broaden the time scale
($\theta_{\rm mas} = 0.015$ mas was used).

\begin{figure}[tbh]
\includegraphics[height=9cm,angle=-90]{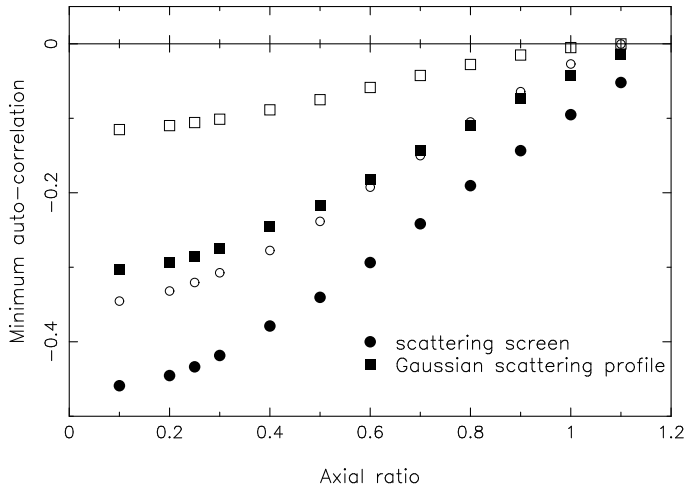}
\caption{Filled symbols show the depth of 
the first minimum in the auto-correlation
of intensity for a point source against the axial ratio of the 
scattering irregularities, for a screen and an extended Gaussian 
profile of scattering strength. The open symbols are for 
a single source whose diameter is large enough to
reduce the scintillation index by a factor of about 3}
\label{fig:overshoot}
\end{figure}

The overshoot phenomenon can be visualized by considering
an intensity pattern in the form of a sinusoidal 
wave such as an idealized ocean wave. 
When sampled along its direction of motion
the time series is a sinewave whose
autocorrelation is a cosine function; this gives
an overshoot depth of -1. 
If a range of wave directions and wavelengths were added,
the first minimum in the correlation would be gradually filled in. 
Thus \it the depth of its first minimum is a useful measure of the 
spectral purity of the phenomenon \rm.
In Figure \ref{fig:overshoot} the filled circles show
how the minimum depth is related to the axial ratio $R$.
As the axial ratio decreases the overshoot saturates with a 
lowest value of -0.48 for a screen.
The curves in Figure \ref{fig:iccfs} give
a satisfactory fit to the observations
and a clear need to include anisotropy, but
it is clear that the parameters are by no means unique. 
Furthermore the 25 pc distance is 20 times smaller than
expected in the TC93 model.
Thus we must explore the trade-off between source size and screen 
distance. First, however, we consider an extended scattering medium as
an alternative to a localized screen.

Under weak ISS the effect of distributed scattering is simply given by 
summing the contributions from each layer in the medium
considered as a screen. (This is sometimes referred to as the 
Born approximation or single scattering, in that the 
perturbations from each layer are assumed to be of low enough
amplitude to neglect multiple scatterings by subsequent layers of an 
already scattered wave).  The correlation function
for an extended scattering medium can thus be found by summing
suitably scaled versions of the screen models weighted according
to an assumed profile of $C_N^2$. We computed the correlation
functions for the Gaussian profile of equation (\ref{eq:cnsq}). 
Compared to the screen curve there are two differences.  
First, the time scale $\tau_I$  for an extended medium of 
scale length, say, 25 pc is only about half of that for a screen at 25 pc.
Second, the negative overshoot is somewhat reduced in amplitude.
This second effect is illustrated in Figure \ref{fig:overshoot}, where
we plot the depth of the first minimum in the acf
as a function of the axial ratio. As before, axial ratios
less than 1.0 correspond to velocity vectors aligned along the
minor axis of the irregularities.  One sees 
that a negative overshoot is obtained for a point source
scattered by anisotropic irregularities in both the screen and 
Gaussian profiles, but that the overshoot is deeper for 
the screen model.   When the source is sufficiently extended
to reduce the scintillation index by a factor 3, the overshoot 
is also reduced, and the reduction is greater for the extended profile. 

In our observations we find overshoot values near -0.55 
with an error of about 0.1, which is just consistent 
with the deepest screen values, and conclude that the axial ratio 
$\simless 0.3$.  Consequently, in 
our detailed modelling of the correlation functions in section 
\ref{sec:iss.stokes}, we emphasize screen models over extended
scattering models and adopt an axial ratio of 0.25, since smaller values
make very little difference. Physically, we imagine that the screen
would be a localized region of relatively dense and irregular
plasma with a relatively uniform mean magnetic
field defining the major axis of the irregularities, 
perhaps covering a quite narrow region of the sky.

The depth of the negative overshoot in the acf can also
be viewed as a measure of the quasi-periodic nature 
of the flux density variations, which are also quantified by
the temporal spectrum for $I$ and  for polarized flux density
$P$, as are shown in Figure \ref{fig:powerspec}.  
The $I$-spectra clearly show
peaks in power near frequencies of 0.3-0.5 cph.
Since the acf is the Fourier transform of the power spectrum,
a peak in the spectrum corresponds to a periodic feature in the acf.
Thus these peaks near 0.3-0.5 cph correspond to periods 
of $\sim 2.5$ hr in the acfs in Figure \ref{fig:iccfs}. 
The overshoots are the minima in the acfs
at half of this period.

Now consider the effect from weak scintillation theory.
The intensity spectrum $P_{II}$ versus spatial wavenumber
is a filtered version of the ISM density spectrum. 
Explicitly (see Coles et al. 1987 and the Appendices of R95)
we have:
\begin{eqnarray}
P_{II}(\kappa_x,\kappa_y) = 8 \pi r_e^2 \lambda^2
C_{N}^2 \delta z \; (R \kappa_x^2 +\kappa_y^2/R)^{-\alpha-2}
\sin^2(z \kappa^2 \lambda/4 \pi) \times \nonumber \\
|V_I(\kappa_x z/2\pi, \kappa_y z/2\pi)|^2
\label{eq:PII}
\end{eqnarray}
Here $r_e$ is the classical electron radius,
the transverse wavenumber is $(\kappa_x,\kappa_y)$
(magnitude $\kappa$). We have used
equation \ref{eq:Pne} for the density spectrum in a layer of thickness
$\delta z$ with $R$ as the projected axial ratio
of the scattering medium and where
$\alpha=5/3$ gives the Kolmogorov spectrum. 

The filter is itself a product of the sine-squared function of 
the Fresnel filter, which cuts off the spectrum at
low wavenumbers, times the squared visibility function 
of the source, which cuts off the spectrum at high wavenumbers.
Thus the filter acts as a bandpass. 
The wavenumber spectrum $P_{II}$ is then mapped to a 
temporal spectrum by the velocity of the pattern relative
to the observer.  Mathematically, this involves a strip-integration
over wavenumbers oriented perpendicular to the velocity vector.
It turns out that, when the velocity 
is oriented within 45 degrees of the wide direction of 
the wavenumber spectrum, there is  a 
significant peak in the temporal 
spectrum displaced from zero frequency as observed 
(Figure \ref{fig:powerspec}). There is no
such peak for the othogonal orientation. As already noted 
such a peak gives a quasi-periodic appearance to 
the flux density variations and a negative overshoot in the acf.

It is important to note that an anisotropic source structure cannot
duplicate this effect.  The negative overshoot in the acf
happens when there is a dip at low wavenumbers in the temporal 
intensity spectrum.  The effect of the source
is to multiply the density spectrum by $|V_{I}|^2$.  Since the 
magnitude of any visibility function must be 
greatest at zero baseline (and equal to
total source flux density), the multiplication
cannot cause a dip at low-frequencies.  However, as the source 
diameter increases it starts to control the scintillation time scale
and if large enough could suppress the oscillatory phenomenon, 
as shown by the open symbols in Figure \ref{fig:overshoot}.

\subsection{Constraining the Scattering Distance, 
Source Diameter and Brightness}
\label{sec:dist.dia}

We now explore the range of possible scattering distances and source
diameters that match the observed scintillation index and time scale.
Here scattering distance $L$ refers either to the distance to a screen 
or to $L_g$ for a Gaussian profile of scattering strength
as in equation \ref{eq:cnsq}.  We computed the acf of intensity 
using weak scintillation theory over a grid of values of
scattering distance and FWHM source diameter. These assumed
either a screen or Gaussian profile of scattering and an axial ratio
of 0.25, as argued above. From each acf the scintillation index
$m_c$ and time scale $\tau_I$ were found by the same algorithm 
as used for the observations.

\begin{figure}[tbh]
\begin{tabular}{lr}
\includegraphics[height=8cm,angle=-90]{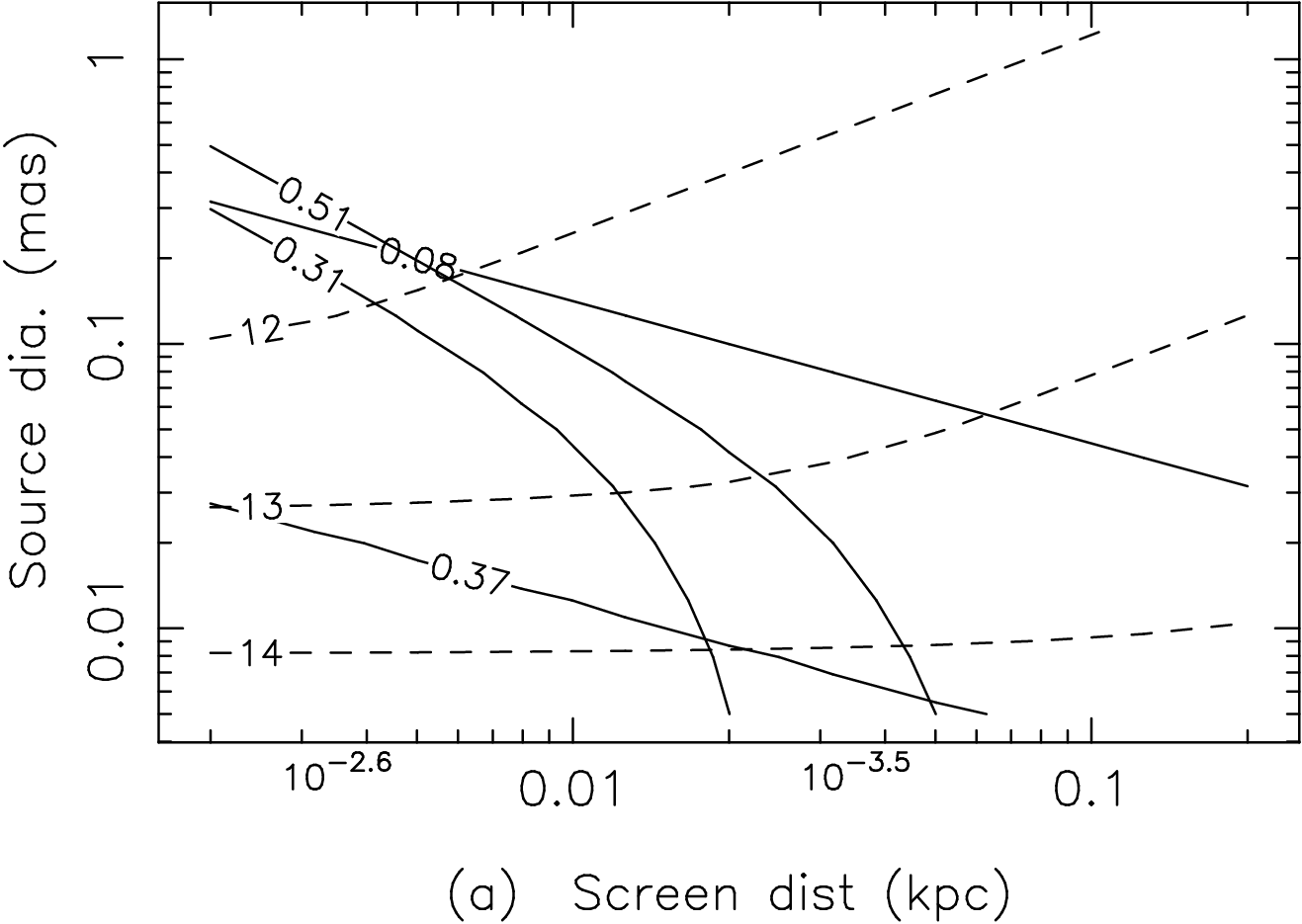} &
\includegraphics[height=8cm,angle=-90]{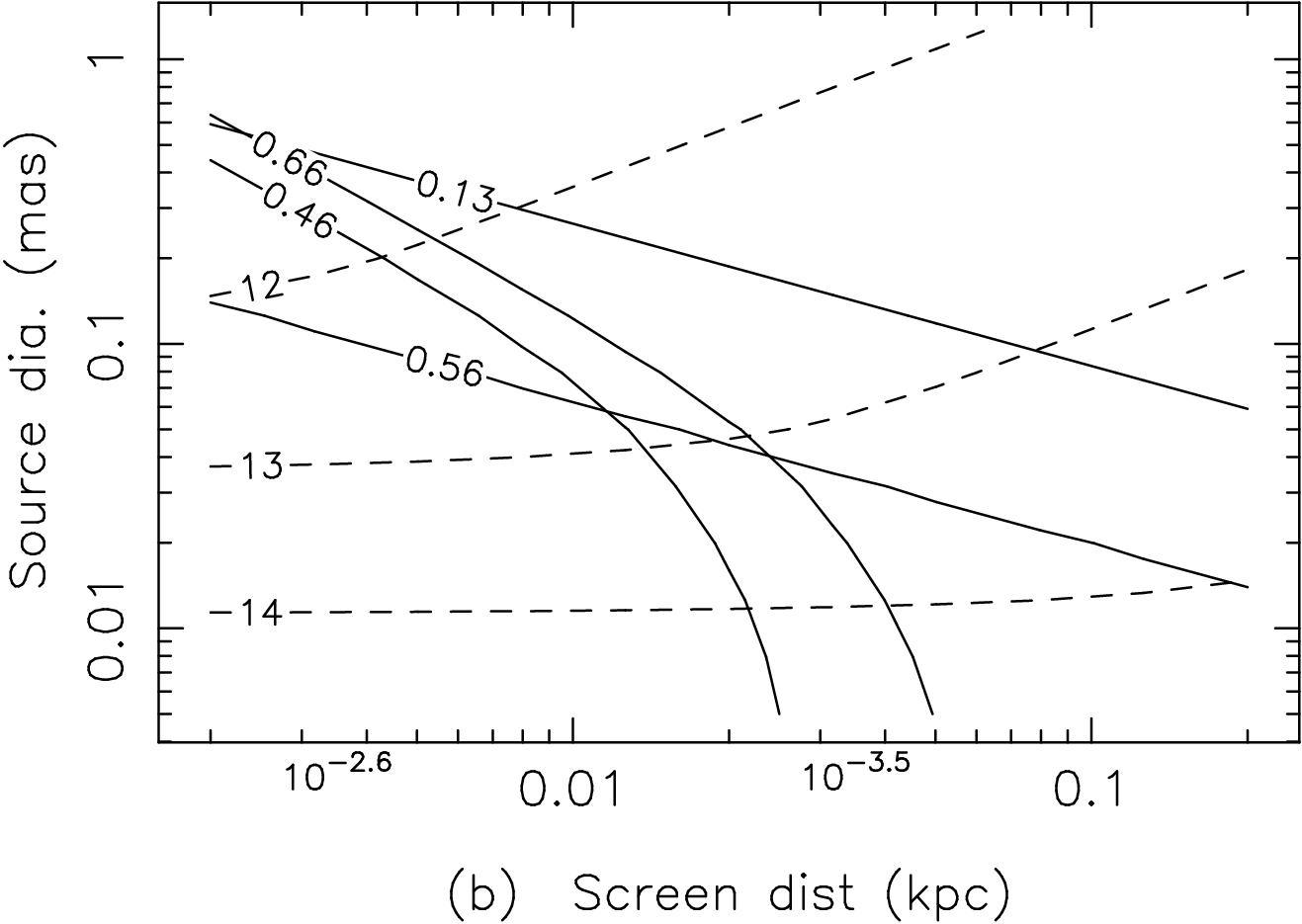} 
\end{tabular}
\caption{The trade-offs possible between scattering distance and
angular diameter of the compact source.  Solid contours are shown for 
the upper and lower observational limits on the scintillation time scale 
($\tau_I$ in hours) and the scintillation 
index $m_c$ for the compact component.
The dashed contours give the source brightness temperature (K) marked
by the exponent of 10.  (a) is for scattering from a screen
at 8.6 GHz  (b) is for a screen at 4.8 GHz.
The distance axis is also annotated with the scattering measure
as explained in the text. 
}
\label{fig:mandtau}
\end{figure}

The results are  shown in Figure \ref{fig:mandtau}a
as contours of $m_c$ and $\tau_I$
in the plane of $\log_{10}(L_{kpc})$, $\log_{10}(\theta_{\rm mas})$
calculated at 8.6 GHz for a screen. 
The observations constrain the values of $m_c$ 
and $\tau_I$, as described in Section \ref{sec:iss.idv}. 
Hence the allowed scattering distance and source diameter
are bounded by the two pairs of solid contours.
Approximately, the screen distance is constrained by
$2 \simless L \simless 50$ pc
with associated source diameter 
$0.3 \simgreat \theta_{\rm mas} \simgreat 0.005$ mas.
The peak brightness of the source can also be estimated 
for each point on the plane and is shown by the dashed contours,
computed as follows. To find the brightness temperature we need
the mean compact flux density $S_c$ and its angular diameter.  The diameter
is given at each point and $S_c$ can be estimated as
$S_{rms}/m_c$ where $S_{rms}$ is the observed rms in flux density and
$m_c$ is calculated for each point.  Thus the diagram shows 
the allowed range in  $\theta_{\rm mas}$, $L_{kpc}$ and $T_b$.
Figure \ref{fig:mandtau}b shows a similar 
screen calculation based on the 4.8 GHz observations.
The latter give the constraints 
$1 \simless L \simless 30$ pc with associated source diameter 
$0.6 \simgreat \theta_{\rm mas} \simgreat 0.04$ mas.

Though the limits on $L$ from the two frequencies
clearly overlap, the two diameter limits at first appear
to be less consistent.  However, in comparing 
source diameters at 4.8 GHz with those at 8.6 GHz, we need to
consider how the effective compact source diameter
might depend on frequency.  We first considered
a uniform self-absorbed synchrotron
source with diameter independent of frequency ($f$).
However, the predicted frequency scaling for $S_{\rm rms}$
was then inconsistent with the observed $S_{\rm rms} \propto f^{-0.5}$.


We proceed by assuming that the core is 
limited by the same maximum brightness temperature
at 4.8 and 8.6 GHz and that $S_c$ has a
spectral index equal to 0.3, as observed for the avearge total
flux density between these frequencies.
This gives the effective diameter 
$\theta_{\rm mas} \propto f^{-1} \sqrt{S_c} \propto f^{-0.85}$.
Accordingly, the angular diameter limits at 4.8 GHz are expected to be
greater by a factor of 1.65 than at 8.6 GHz, though the factor is
uncertain because of uncertainty in the fraction of flux density
in the compact core. So in Figure \ref{fig:mandtau}b 
the 4.8 GHz angular diameter divided by 1.65
can be interpreted as the effective diameter at 8.6 GHz 
and compared with limits in Figure \ref{fig:mandtau}a.
Thus dividing the 4.8 GHz diameter limits by 1.65 we find
$0.36 \simgreat \theta_{8.6, \rm mas} \simgreat 0.025$ mas.
The joint constraint 
on the diameter at 8.6 GHz
is $0.36 \simgreat \theta_{8.6, \rm mas} \simgreat 0.025$ mas
associated with $2 \simless L \simless 30$pc and
$5 \times 10^{11} \simless T_b \simless 2 \times 10^{13}$ K. 
Also noted on the horizontal axis of
the diagram are values for the scattering measure of the screen
($SM$), which is the line-of-sight integral of $C_N^2$.

The allowed ranges of $L$ and $\theta_{\rm mas}$ 
also depend on several of the 
other assumptions in the model which we now discuss.
Changing the axial ratio to 1.0 causes longer scintillation times
and so shifts the time scale contours about a factor 1.4 toward
even smaller distances; however, we have already 
argued that isotropic models are excluded, 
since they do not explain the negative overshoot in the acfs. 

Another variable is the transition frequency $f_w$ between weak and 
strong scintillation, which we have taken to be 4.0 GHz.
For a point source $m_c$ peaks near $f_w$.
As the source size increases $m_c$ is reduced and
its peak is shifted to frequencies below $f_w$,
as can be seen from Figure 3 of R95. 
We observe a peak in $m_c$ near 4.8 GHz
and so $f_w$ could possibly be above 4.8 GHz.
Such a change would shift the $m_c$ contours upward in
Figure \ref{fig:mandtau}, which further reduces both
the scattering distance and the source brightness.


We also examined the effect of a 
Gaussian profile of scattering with the 1/e path length $L_g$.
Since such a model includes some scattering very close to
the Earth, this allows an increase in $L_g$ (compared to screen distance $L$),
by a factor of about 1.5, allowing a smaller diameter 
(and brighter) source and a scattering path length up to 70 pc.  
However, given that the Gaussian profile
does not fit the overshoot behavior in the auto-correlations,
we now confine our attention to the screen models.

 
The velocity for the diffraction pattern is another important 
parameter, which we assume to be 36 km s$^{-1}$,
the velocity of the Earth relative to the LSR. 
Increasing the velocity to 75 km s$^{-1}$
(the upper limit from Jauncey et al., 2000)
changes the mapping from time scale to spatial scale,
allowing a larger distance and smaller source size.
The contours for time $\tau_I$ depend on $V$, $L$ and
$\theta_{\rm mas}$ 
such that the upper limit for $L$ 
scales as $V_{\rm iss}^2$, while the lower limit scales
linearly with $V_{\rm iss}$. The combined limits from 
plots as in Figure \ref{fig:mandtau} recomputed for 
75 km s$^{-1}$ give
ranges  $8 \simless L \simless 100$ pc associated with 
$0.13 \simgreat \theta_{8.6, \rm mas} \simgreat 0.013$ mas
and $2 \times 10^{12} \simless T_b \simless 6 \times 10^{13}$ K.

The conclusion from the foregoing, rather lengthy discussion is that
for our preferred model (with $f_w \sim 4$ GHz) the modulation 
index and time scale for IDV of PKS 0405-385 at 4.8 and 8.6 GHz
require a very local scattering region, in a screen
in the range 2 to 30 pc.  The overshoot in the acfs, which
require substantially anisotropic scattering 
(axial ratio $\simless 0.25$).  We presume that
the orientation of the anisotropic structure in the plasma is
determined by the local mean magnetic field and that
a pronounced anisotropy implies a uniform field through
the scattering plasma. This is more reasonable physically 
if the scattering is localized in a thin layer
(screen) rather than in an extended scattering medium, 
in which the magnetic field would tend to be disorganized.

Associated with the distance and velocity limits there are allowed ranges of
maximum source brightness in the range
$5 \times 10^{11}$K to $\sim 6 \times 10^{13}$K for the nominal
screen model.  Our nominal brightness limits are 
about 25 times lower than inferred by KCJ,
who based their analysis on the greater scattering distances of
the TC93 model for the IISM.  
We also note that there is an inverse 
relation of brightness and the fraction of the total flux density 
in the compact core (100\% for the lowest
brightness down to about 25\% for the highest brightness). 

\section{ISS of the Stokes Parameters}
\label{sec:iss.stokes}

In several IDV sources the fluctuations in $I$ are 
accompanied by fluctuations
in the linear polarization parameters $Q$ and $U$.
Figures \ref{fig:xIQU} -- \ref{fig:lIQU} ,  \ref{fig:xc.IQU.cor} \&
\ref{fig:sl.IQU.cor} show that, for PKS~0405-385 the
polarization parameters vary faster than does $I$, 
as is often the case for other IDV sources (Q89, Gabuzda, 2000a,b,c). 
As discussed below, in the ISS of a linearly polarized 
\it point-source \rm $Q(t)$ and $U(t)$ should simply be 
scaled versions of the variations in $I(t)$. 
Figures \ref{fig:xIQU}, \ref{fig:cIQU}, \ref{fig:sIQU} 
and \ref{fig:lIQU} show that the Stokes' variations
in PKS 0405-385 are clearly inconsistent with such behaviour.

As will be shown below, when the polarization structure
differs from that of the total flux density, the $Q$ and $U$
scintillations become partially decorrelated from those 
in $I$.  We now develop the appropriate theory and 
in section \ref{sec:fitting} we search for
a particular polarized source structure
that matches the observations of PKS 0405-385.
Though observers often display fluctuations in polarized flux density 
$P=\sqrt{Q^2+U^2}$ (or polarized degree $p=P/I$) 
and position angle $\chi=0.5 \tan^{-1}(U/Q)$,
we choose $Q,U$, since they
sum linearly from incoherent regions of emission.

We analyze ISS from a source whose structure in $Q$ and $U$ may differ
from that in $I$. An entirely parallel formulation can be done for $V$,
though it is not considered in this paper.
A brief outline of these results was given by Rickett (2000) and an
entirely similar analysis was used by Medvedev (2000) in the context of 
strong scintillations, applied to the radio afterglows of gamma ray 
bursts. In Section \ref{sec:analysis} 
we  quantified the relationship between
fluctuations in $I$,$Q$ and $U$ by their six auto- and cross-correlation
functions. Thus we develop the theory for these correlations,
assuming weak or refractive ISS.  
The method is simply a repeated application
of the smoothing formula as expressed  
by Little and Hewish (1966) and Salpeter (1967).

Consider a scattering layer (screen) of thickness $\delta z$
at distance $z$ from an observer who receives polarized waves from
a source whose brightness distribution in $I$ is
$B_I(\theta_x,\theta_y)$.  Let $\Delta I_o(x,y)$ be
the pattern of intensity fluctuation about its mean for a unit
flux density point source located on the z-axis
at a great distance.  The 
net flux density is the sum of shifted terms weighted by each
brightness element:
\be
\Delta I(x,y) = \int \int  \Delta I_o(x-z\theta_x,y-z\theta_y) 
B_I(\theta_x,\theta_y) d\theta_x d\theta_y  \; ,
\label{eq:dI}
\ee
where infinite limits are implied in this and subsequent integrals
since the brightness functions are confined to small angles.
Exactly similar equations can be written for 
$\Delta Q$ and $\Delta U$ for a source with
polarized brightness functions  $B_Q(\theta_x,\theta_y)$ and 
$B_U(\theta_x,\theta_y)$, since the ISS in, say,  $Q$
due to each source element is simply 
$\Delta I_o(x-z\theta_x,y-z\theta_y) 
B_Q(\theta_x,\theta_y) d\theta_x d\theta_y $.

The important underlying assumption
here is that the scintillation conditions in the
right and left circular modes of propagation 
through the magneto-ionic plasma are essentially identical.  
For radio-astronomical signals travelling through the ISM
the phase difference in the two modes is
twice the angle of Faraday rotation on that line of sight.
From the mean polarization angle observed versus wavelength 
we estimate a rotation measure $RM \sim 70$ rad m$^{-2}$, 
which at 8.6 GHz  gives $\simless 0.2$ radian 
phase difference between left and right. 
This is very small compared to the total phase increment due 
to the plasma, for which a typical value might be 
$10^7$ radians ($\propto \lambda$).  
To put it more precisely, any scintillation 
in $p$ or $\chi$ for a point source will be negligible, 
if across a Fresnel scale there is a negligible
\it change \rm in this right-left phase difference.
Observational evidence for this condition 
in the typical ISM is obtained from
Haverkorn et al. (2000), who used the Westerbork
telescope to observe the polarized Galactic background. 
They observed about 1 radian changes in its polarization angle
over a scale as small a 4 arc-minutes near a Galactic latitude 
of 16 degrees at a wavelength of 1.1 meter. 
At an assumed distance of 500 pc for the magneto-ionic
medium, this corresponds to about 0.5 pc 
and maps at 8.6 GHz to a typical change in the 
Faraday phase on scale $r_F$
of about $10^{-8}$ radians . This 
confirms that there should be negligible 
point-source scintillation in the degree or 
angle of polarization.

\begin{figure}[htb]
\begin{tabular}{lr}
\includegraphics[height=7.5cm,angle=-90]{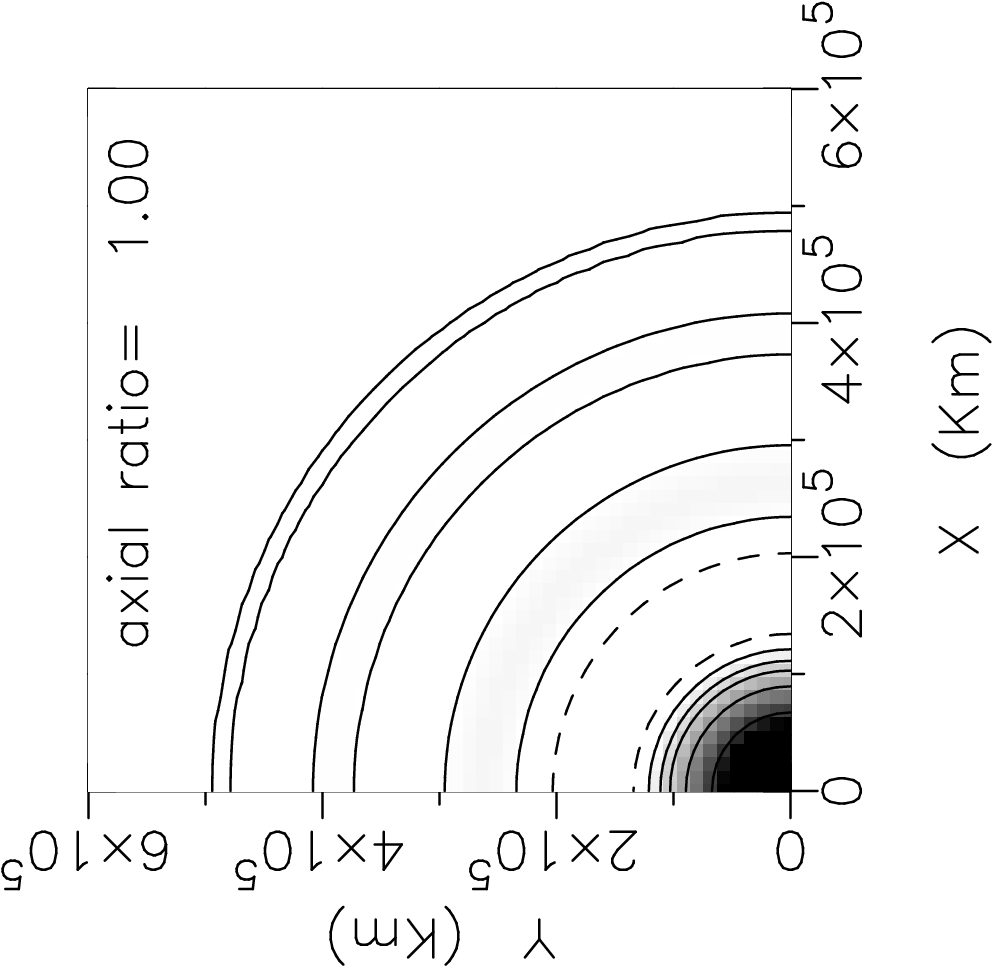} &
\includegraphics[height=7.5cm,angle=-90]{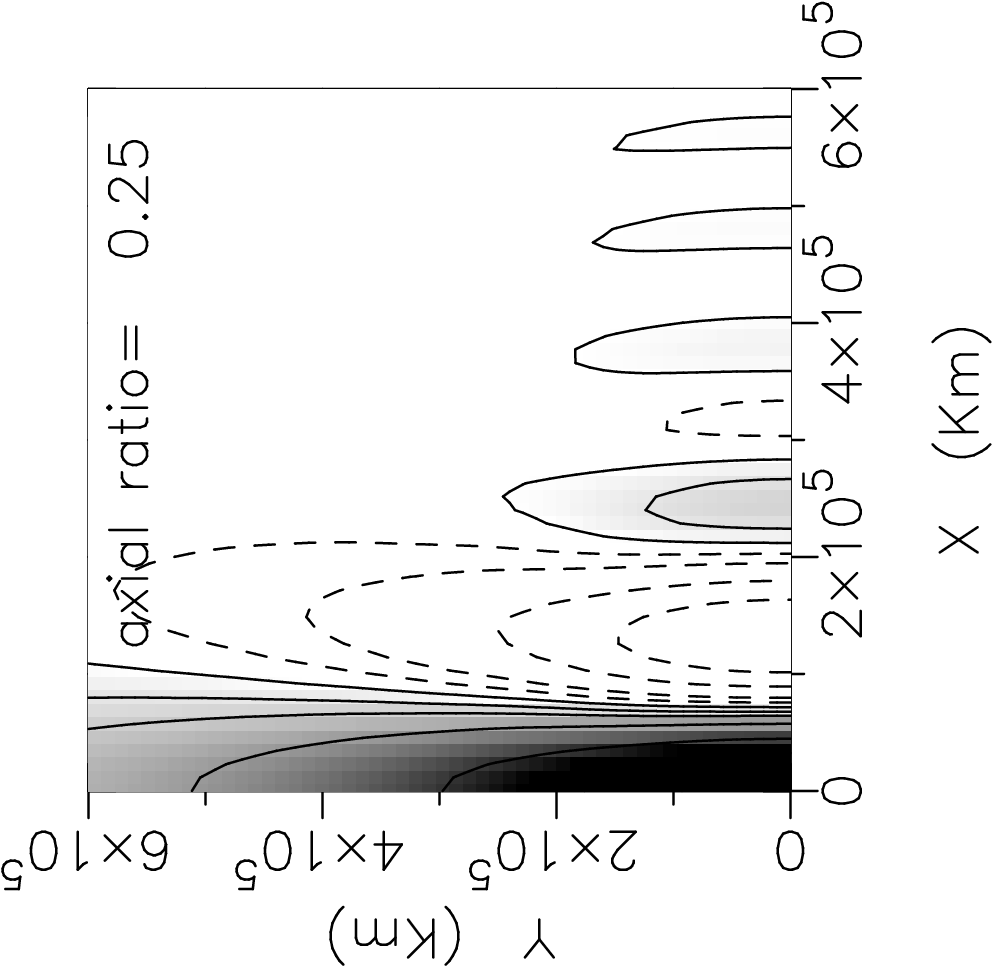} \\
\end{tabular}
\caption{Contour plots of $C_{oo}(\vec{\sigma})$
in a single quadrant (function is symmetric about both axes).
Axial ratio 1.0 in left panel and 0.25 in right panel.
The dashed contours are negative at (-0.05, -0.1, -0.2, -0.3)
only for only for the latter are there deep negative lobes; 
the shaded solid contours visible are at (0.4, 0.2, 0.1, 0.05, 0.0). 
Here $r_F = 6.7\times 10^4$km.}
\label{fig:ptspread}
\end{figure}

For a very distant point source the ISS variations due to
a screen are approximately statistically homogeneous, 
thus the spatial auto-correlation for $\Delta I_o$
depends only on the spatial separation $(\xi,\eta$), not 
on absolute position:
\be
C_{oo}(\xi,\eta) = <  \Delta I_o(x,y)   \Delta I_o(x+\xi,y+\eta) >
\label{eq:Coo}
\ee
Similar correlations versus spatial offsets can be defined for
each pair of observed Stokes' parameters in a fashion analogous to
equations (\ref{eq:Ctdefs}), for example:
\be
C_{IQ}(\xi,\eta) = < \Delta I(x,y) \Delta Q(x+\xi,y+\eta) > 
\label{eq:Cxydefs}
\ee
Substituting equation (\ref{eq:dI}) and the parallel equation for 
$\Delta Q$ into (\ref{eq:Cxydefs}), one finds:
\be
C_{IQ}(\xi,\eta) = \int  \int  C_{oo}(\xi-z \alpha_x,\eta-z \alpha_y)
R_{IQ}(\alpha_x,\alpha_y) d \alpha_x d \alpha_y  \; ,
\label{eq:Ciq}
\ee
where
\be
R_{IQ}(\alpha_x,\alpha_y) = \int  \int B_I(\beta_x-\alpha_x/2,\beta_y-\alpha_y/2) 
B_Q(\beta_x+\alpha_x/2,\beta_y+\alpha_y/2) d\beta_x d\beta_y \; .
\label{eq:RIQ}
\ee
There are also similar results for the other
\sloppy 
auto- and cross-correlations ($II$, $QQ$, $UU$, $IU$, $QU$).  
\fussy
Each gives the observed correlation function as a convolution of
the cross-correlation of the two brightness distributions
by an ISS resolution function $C_{oo}(\xi,\eta)$, which
is plotted in Figure \ref{fig:ptspread} for axial ratios 
1.0 and 0.25.

As mentioned in Section \ref{sec:iss} we are 
concerned with refractive ISS in strong scintillation or with
weak ISS.  In both cases the convolution results apply
and so theoretical expressions for
$C_{oo}(\xi,\eta)$ are given by products in the wavenumber domain. 
Thus each spectrum is the product of the point source scintillation 
power spectrum by a source filter, which is the Fourier transform of the source 
brightness correlations $R_{II}$, $R_{IQ}$ etc. 
For $R_{II}$ the source filter is the squared magnitude of the visibility 
function as already given explicitly in equation \ref{eq:PII}. 
This is the so-called ``Cohen-Salpeter'' equation (Salpeter, 1967).  
For the cross-spectrum between $I$ and $Q$
the filter becomes complex (e.g.\ $V_I V_Q^*$), giving:
\begin{eqnarray}
P_{IQ}(\kappa_x,\kappa_y) = 8 \pi r_e^2 \lambda^2
C_{N}^2 \delta z \; (R \kappa_x^2 +\kappa_y^2/R)^{-\alpha-2}
\sin^2(z \kappa^2 \lambda/4 \pi) \times \nonumber \\
V_I(\kappa_x z/2\pi, \kappa_y z/2\pi) V_Q^*(\kappa_x z/2\pi, \kappa_y z/2\pi) 
\label{eq:PIQ}
\end{eqnarray}
Note that the x-direction is defined here by the orientation
of the ellipse in wavenumber, such that for $R<1$
the minor axis of the spatial correlation ellipse lies along
the x-axis; this defines the coordinate frame for 
our fitted source models. 
Equations (\ref{eq:PIQ} and \ref{eq:PII}) apply to 
\it weak \rm scintillation and to refractive
scintillation in which the source filter cuts-off the scintillations
at a lower wavenumber than the refractive cut-off (at
the inverse of the scattering disc size). Such conditions seem to apply
for most extragalactic sources observed away from the Galactic plane.

We compare theory with the observations in the time domain and so
we compute the model as a two-dimensional Fourier transform 
of each wavenumber spectrum -- for example:
\be
C_{IQ}(\xi,\eta) = \int \int P_{IQ}(\kappa_x,\kappa_y) 
\exp[i(\kappa_x \xi+ \kappa_y \eta)] d\kappa_x d\kappa_y .
\label{eq:CIQborn}
\ee
We then map spatial lags to temporal lags for a given model velocity.

The other geometry to be considered is a scattering medium that
extends from the observer out through the Galactic disc.  We have
chosen to model this by the Gaussian in equation (\ref{eq:cnsq}) 
with scale length $L_g$.
Putting this into equation (\ref{eq:PIQ}) and integrating over $z$
from zero to infinity gives expressions for the auto- and cross spectra,
which can be integrated analytically for Gaussian brightness
functions.  Details are shown in Appendix B.
The Gaussian functions give explicit equations
that allow efficient computation.  It is unlikely that different functional
forms for the tapering of the tail of profile will make 
much difference to the calculations. Many other geometries can be considered
between the extremes of a screen and a Gaussian profile.
Coles (1988) shows that, when the scattering
medium does not continue up to the observer for even a small 
fraction of $L_g$, the shape of the acf starts to approximate
the screen result.

\section{Model-Fitting the Stokes' Correlations at 8.6 GHz}
\label{sec:fitting}

We note that it is not possible in general 
to construct model time series
for $Q(t)$ and $U(t)$ from $I(t)$ for more than one 
source component, since the ISS creates 
a two dimensional stochastic pattern, of
which the observations only sample a one-dimensional slice.
Consequently, we compare the 8.6 GHz Stokes' correlations
with computed models using the theory of the previous section. 
We initially considered an algorithm
like CLEAN, as used in image synthesis in which a
model for the visibility function is built
by adding a sequence of Gaussian components.
However, this is not feasible here, since 
the Stokes' correlations depend on the square of the
visibility (equation \ref{eq:PIQ}).
The associated cross-products between every pair of 
components makes the addition of components non-linear. 
Thus we used conventional least-squares minimization
on the model correlation functions.

The models require specification of both the scattering medium and 
the polarized source structure, with parameters summarized
in Table \ref{tab:params}.
The important parameters of the scattering medium have
already been discussed. In our fitting we fixed $L=25$ pc, 
$\alpha = 5/3$, at the Kolmogorov value.  
We also fixed the axial ratio $R$ to be 0.25 
and the velocity of Earth relative to the scattering plasma $V_{iss}$
as described in section \ref{sec:iss.idv}; though its orientation
relative to the minor axis of the scattering irregularities
was a fitted parameter.
 
We model the source structure as the superposition of circular
uniformly polarized Gaussian components. 
Each is characterized by its flux density, its
fractional linear polarization and position angle, 
its FWHM diameter and its
angular position relative to a central component,
again referred to the minor axis of the scattering irregularities. 
The number of parameters to model 
the source increases from 4 for a single source 
to 10 for two components to 16 for three components.  
We now describe attempts to model the 
data with one, two and three components.


We first considered a single Gaussian component
of uniform polarized fraction $p_1$ and position angle $\chi_1$.
In this case the brightness distributions
of $Q$ and $U$ are exactly proportional
to the brightness distribution in $I$, and so
the polarized ISS will be like that for a polarized point source,
in that the fluctuations $Q(t)$ and $U(t)$ should 
be simple linearly scaled
versions of $\Delta I(t) = I(t)-<I>$.
This model requires four constant coefficients, which we
estimated from a least-squares fit to the observed $Q(t)$ 
and $U(t)$, as shown by the solid lines in Figure \ref{fig:xIQU}.
These are clearly unsatisfactory.   Expressed another way,
the six Stokes' correlation functions of
Figure \ref{fig:xc.IQU.cor}a would all follow the
form of the acf, except that the ccfs
could be sign inverted depending on the signs of $Q$ and $U$.
This is also clearly incompatible with the observations. 
We conclude that a single component of uniform 
polarized fraction and position angle
does not explain the polarized IDV.
We note that this analysis and conclusion apply in the presence of 
an added non-scintillating component with a different 
linear polarization, even though in such a case
the overall polarized fraction and angle may become time variable. 


\subsection{Multi-Component Models}

Moving to multi-component source models, we use a non-linear least-squares
optimization scheme\footnote{
The optimization used the code ``DNLAFU'' distributed   
by C. L. Lawson at JPL, based on ``NL2SOL'' originated
by Dennis, Gay, and Welsch, 1981 (with developments by Gay and 
Kaufman at AT\&T Bell Labs).}
to search for source parameters to best fit
the six Stokes' correlations. 
In Figure \ref{fig:xc.IQU.cor} we plotted the correlations averaged
over the three full 12-hour tracks of the source (June 8, 9, 10)
at both 4.8 and 8.6 GHz.  It is difficult to
estimate the statistical errors associated with these functions, and so
we also computed the correlations for each of the
individual days.  Here it was clear that the main features of the 8.6 GHz
correlations were recognizable in both auto- and cross-
correlations for each of the individual days; however, at 4.8 GHz the features
in the cross-correlations were much less repeatable.  Consequently, we
concentrate on model-fitting to the 8.6 GHz observations, 
since they show the most consistent 
behaviour in the correlation functions.

The theoretical expressions for 
the temporal correlations are obtained
from the spatial correlation functions  
by mapping spatial lags to temporal lag ($\tau$) 
according the the velocity model
($V_{iss,x},V_{iss,y}$).
\be
C_{IQ,t}(\tau) = C_{IQ}(\xi=V_{iss,x}\tau, \eta=V_{iss,y}\tau) .
\label{eq:ciq}
\ee
Here $C_{IQ}(\xi,\eta)$ is given by the Fourier transform
(\ref{eq:CIQborn}) of the spectrum from equation (\ref{eq:PIQ}).
Thus for a given set of model parameters we compute the six
\sloppy
correlations $C_{II,t}$, $C_{QQ,t}$, $C_{UU,t}$, $C_{IQ,t}$, $C_{IU,t}$, $C_{QU,t}$
\fussy
at the time lags observed.  We minimize the following 
function, which is the weighted sum of 
the squared differences (observation - theory) 
from all six correlation functions:
\be
S^2 = \Sigma^{i,q,j} wt(\tau_j) 
[ C_{iq,t,obs}(\tau_j) - C_{iq,t,theory}(\tau_j) ]^2/\sigma_{iq,j}^2 / 
\Sigma^{i,q,j} wt(\tau_j)
\ee
where $i$ and $q$ are each $I,Q$ or $U$, $\sigma_{iq,j}$ is the rms error
in the correlations and $ wt(\tau_j)$ is a 
triangular weight falling to zero at times of $\pm3$ hrs. 
We choose these limits on time lag since it is known
that errors in correlation estimates remain high at large lags,
whereas the signal is expected to decrease.  Therefore we
do not consider source models in which there could be components
separated by angles substantially greater than their diameters, 
since in that case correlations could extend to large time lags.

\begin{figure}[tbh]
\includegraphics[height=10cm,angle=-90]{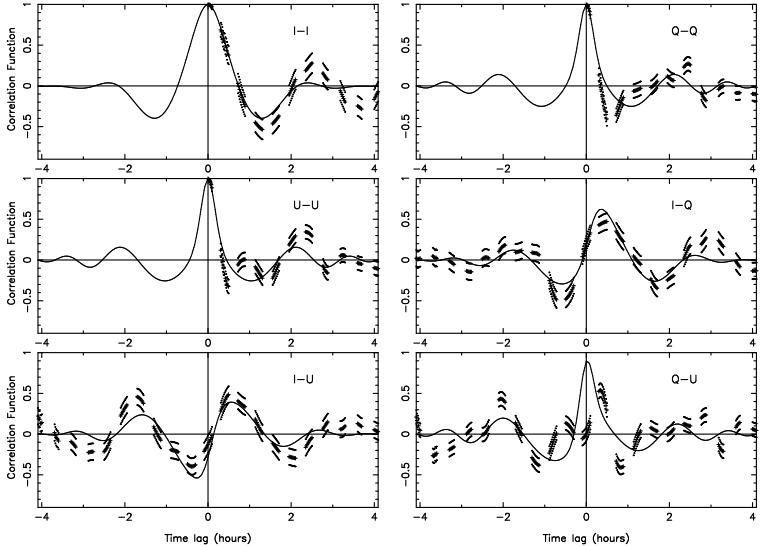} 
\caption{Auto- and cross-correlations at 8.6 GHz as in 
Figure \ref{fig:xc.IQU.cor}, overplotted with a 
best fit model with 2 scintillating source components and
scattering localized in a screen.  }
\label{fig:2compx}
\end{figure}

\begin{figure}[tbh]
\includegraphics[height=10cm,angle=-90]{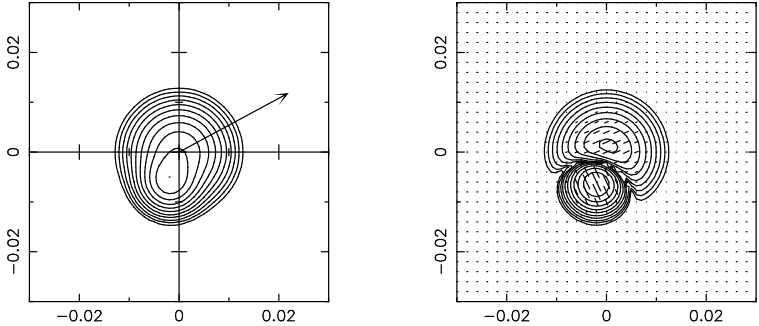}
\caption{ Brightness distributions for the two component source 
model used in Figure \ref{fig:2compx}.  
The angular scale is in mas, with the horizontal axis
parallel to the minor axis of the ISM anisotropy ellipse
(whose absolute orientation is not known).
The relative direction of velocity vector is shown by the arrow.
The left panel shows the total brightness temperature with the peak at
$\sim 10^{14}$K.  The right panel shows contours of polarized brightness 
$P$  with a maximum of 53\% of the total with 
bars showing $P$ and position angle $\chi$.
The contour levels decrease logarithmically (5 levels per decade)}
\label{fig:2comp.polb}
\end{figure}

With two scintillating source components, there are 15 parameters, 
of which 4 were fixed at the values given in Table \ref{tab:params}.
In searching the remaining 11 parameters we find
a satisfactory fit to five of the six correlation functions, 
as shown in Figure \ref{fig:2compx}.
However, the model fails to reproduce the offset peak
visible at a time lag of 0.3 hrs in $C_{QU,t}(\tau)$. 
Further it does not fit the $Q-Q$ and $U-U$ correlations
very well. The minimum value of the reduced chi-squared
error ($S^2$) is 5.6, significantly worse than 1.0 expected for
a satisfactory fit.  If we exclude
$C_{QU,t}(\tau)$ a better fit can be obtained to the
remaining five correlations with a (satisfactory) minimum $S^2 \sim 1.4$.
Figure \ref{fig:2comp.polb} shows the source
brightness distribution for the fit of Figure \ref{fig:2compx}.
The brightness plots were calibrated in Kelvin by the following
procedure.

The scaling of the brightness contours depends on the 
total compact flux density ($S_c$), which is
not estimated during the model fitting, since the auto-
and cross-correlations are normalized. However, 
$S_c$ can be estimated from $S_{rms}/m$, where
$m$ is the scintillation index from the model.  In addition to the 
other parameters, this requires an estimate of the 
scattering measure (SM), which we determined from
our constraint on the transition frequency 
$3.0 \simless f_w \simless 4.8$ GHz, as already discussed;  
the corresponding uncertainty in the maximum derived 
brightness is about a factor two.

The brightness has a maximum of close to $10^{14}$K at the origin
of the plot and the degree of polarization is a maximum of
43\%.  These interesting physical estimates will be 
discussed in the next section after presentation of
three-component models.  The basic topology that 
fits the polarization behaviour is
a pair of components with approximately orthogonal linear polarization,
separated by about 10 $\mu$as along the ISS velocity direction,
which is inclined at about 30 degrees to the minor axis 
of the scattering irregularities.
The components must also be displaced by some distance
perpendicular to the velocity since the maximum $QU$ cross
correlation is not 100\%; a perpendicular
separation of 5-10 $\mu$as ensures that the scintillation
from the two components are only partially correlated.
Since we were unable to find a two component model that
matches all six correlations and we believe that the offset in
the peak in $C_{QU,t}(\tau)$ is significant, we tried three components.

\begin{figure}[tbh]
\includegraphics[height=12cm,angle=-90]{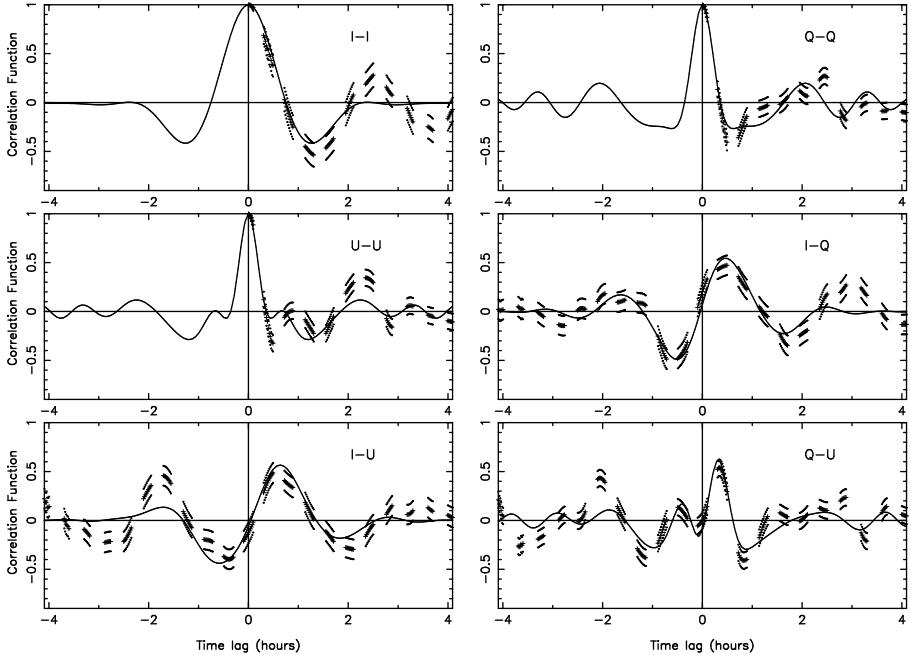}
\caption{Auto- and cross-correlations at 8.6 GHz as in 
Figure \ref{fig:xc.IQU.cor}, overplotted 
with a fit to a 3-component model. This fit minimizes the
peak brightness temperature among the other similar models}
\label{fig:3compx}
\end{figure}

\begin{figure}[tbh]
\includegraphics[height=10cm,angle=-90]{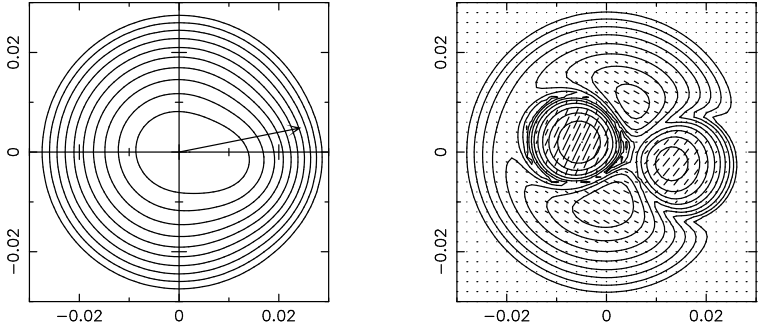}
\caption{ Brightness distributions for the three component source model
used in Figure \ref{fig:3compx}.  The angular scale is in mas plotted
in the format of Figure \ref{fig:2comp.polb}. The left panel is the 
total brightness temperature with the peak 
at about $2 \times 10^{13}$K.  The right panel is the polarized
brightness with maximum  $\sim 70$\% of the total brightness.
}
\label{fig:3comp.polb}
\end{figure}

With three scintillating source components, there are 21 parameters, 
of which 4 were fixed as given in Table \ref{tab:params}.
In searching the 17 variable parameters we find good
fits to all six correlation functions.  However, the optimum
is not uniquely determined -- there is a set of satisfactory solutions, 
with minimum $S^2$ of about 1.4.  Since there is still 
substantial uncertainty about the correct normalization of the errors
in the auto- and cross-correlations, there is a corresponding 
uncertainty about the absolute normalization in $S^2$.
Consequently, we accept 1.4 as a satisfactory fit and do not attempt
a more complex model, which would require even more parameters.
The example shown in Figure \ref{fig:3compx} is one which minimizes
the peak brightness temperature.  
The brightness distribution in total flux density
and in polarized flux density are shown in Figure \ref{fig:3comp.polb},
in which the maximum total brightness temperature is $\sim 2\times 10^{13}$K
and the peak polarization degree of $\sim 70$\%.
Note that the model uses three components of uniform polarization
degree and angle, with the central component extending over $\sim 30 \mu$as.
Since its polarization is nearly orthogonal to that of the other
two smaller components, the net polarization has a minimum
near the center, creating four regions of polarized brightness.

There are in fact many other source models with
higher peak brightness (but no significant improvement in the fit), but
there are no satisfactory fits with significantly lower brightness temperatures
than in the fit displayed. Thus though the model shown here 
is by no means unique, it clearly
demonstrates that a source model can be found to
give a quantitative fit to the correlations among the Stokes' parameters
with a physically reasonable degree of polarization and with
a maximum brightness $\sim 2\times 10^{13}$K. While this is high
it is consistent with Doppler beaming factors implied from 
VLBI measurements of apparently superluminal motion in other AGNs.

\begin{table}[htb]
\begin{tabular}{cccc}
\tableline
Symbol & Parameter & 2-comp & 3-comp \\
\tableline
$L$ & screen distance (pc) & 25* &  25* \\
$R$ &  axial ratio &  0.25* & 0.25*   \\
$V_{iss}$ &  ISS velocity (km s$^{-1}$) &  36* & 36*   \\
$\psi$ &  direction of ISS velocity &  $28^{\circ}$ & $11^{\circ}$   \\
$\alpha$ &  spectral exponent &  5/3* & 5/3*   \\
$S_{I,1}$ & Total flux density (Jy) &  0.38  & 0.48  \\
$\theta_1$ & FWHM ($\mu$as) &  15 & 30  \\
$P_1$ &  polarized flux density (Jy) &  0.18  & 0.33 \\
$\chi_1$ &  polarized PA &  $23^{\circ}$ & $-29^{\circ}$  \\
$S_{I,2}$ & Total flux density comp 2 (Jy) &  0.19  & 0.024   \\
$\theta_2$ & FWHM ($\mu$as)  &  9.4 & 19  \\
$P_2$ &  polarized flux density comp 2 (Jy) &  0.17  & 0.22    \\
$\chi_2$ &  polarized PA  &  $-65^{\circ}$ & $63^{\circ}$    \\
$\theta_{ox,2}$ & comp 2 position ($\mu$as)  & -1.9   & -4. \\
$\theta_{oy,2}$ & comp 2 position ($\mu$as) & -6 & 1.6 \\
$S_{I,3}$ & Total flux density (Jy) &  0 &  0.091  \\
$\theta_3$ & FWHM ($\mu$as)  & - & $19\mu$as  \\
$P_3$ &  polarized flux density comp 2 (Jy) & - & 0.11 \\
$\chi_3$ &  polarized PA  & - & $56^{\circ}$    \\
$\theta_{ox,3}$ & comp 3 position ($\mu$as)   & - & 11 \\
$\theta_{oy,3}$ & comp 3 position ($\mu$as)    & -  & -2 \\
\tableline
\end{tabular}
\caption{Model parameters. The values with an asterisk were fixed in the fits.
Angles are given relative to the minor axis of the scattering ellipse
(defined in equation \ref{eq:PIQ}).
Polarized position angles are defined as for the observations.
Though the polarized flux densities for components 2 and 3 are
greater than 70\% of their individual total flux densities, the polarized 
brightness is everywhere less than 70\% of the total
brightness in these models, as displayed in 
Figures \ref{fig:2comp.polb} and \ref{fig:3comp.polb}.
See text concerning the lack of uniqueness in the parameters.
}
\label{tab:params}
\end{table}

The fits displayed and discussed above are all based on a 
screen model for the scattering medium.  As discussed in
section \ref{sec:aniso} the effect of a Gaussian profile
of scattering was also computed and shown to reduce
the amplitude of the overshoot in the auto-correlations.
While the screen model gives less overshoot than observed,
it does agree within about $1\sigma$.  The
Gaussian profile models gave less overshoot and disagree
by $\simgreat 2\sigma$. In view of this systematically worse
fit, we did not attempt source model fits for the Gaussian profile.

\subsection{Stokes' Correlations at 4.8 2.4 and 1.3 GHz}
\label{sec:CSL}

The Stokes' parameters at the three lower frequencies have
generally lower amplitudes than at 8.6 GHz (Table \ref{tab:rms.tau})
and their correlations have greater
statistical errors.  Consequently, we did not attempt detailed
model fits to their Stokes' correlation functions.
Nevertheless, the lower frequency results are part of the IDV
data to be explained.  There are two relevant effects.  
The first is the possible suppression
of ISS in the local screen by angular broadening in the ``normal''
Galactic plasma scattering. The second is increasing Faraday rotation and
depolarization in the source itself.  We suggest that 
a combination of these effects explains the lower 
amplitude IDV in $Q$ and $U$ and the lack of systematic correlation
with the IDV in $I$.


The TC93 model predicts a path length through the IISM
of about 1000 pc toward PKS 0405-385, at a Galactic latitude of -48 degrees.
The model is a smooth representation of the dispersing and scattering
observed for pulsars, which is known to be quite ``patchy''.
However, it would be surprising if on our line of sight there
were no more distant scattering electrons than at 25-100 pc.
The scattering measure ($SM$) on that line of sight, found
by integrating the TC93 model for $C_N^2$, is
$SM = 2.0 \times 10^{-4}$ m$^{-20/3}$ kpc, which 
is comparable to the values for the scattering model
used in our model and in Figure \ref{fig:mandtau}. 

This must cause an angular broadening, which 
may be sufficient to suppress the
scintillation in the local region.  The  
angular scattering diameter (FWHM), 
which depends on $SM$ and the frequency
(see Cordes et al., 1991, for example), 
is 7 $\mu$as at 8.6 GHz and 24 $\mu$as at 4.8 GHz.  
This will be most critical in the ISS of the smallest structures
in our 8.6 GHz model. which are low flux density highly polarized
components with diameters $\sim 20 \mu$as. If at 4.8 GHz
these were scatter broadened to 24 $\mu$as the
rms amplitude of their fluctuations would 
be significantly decreased over those at 8.6 GHz,
while at the same time the rms amplitude in 
$I$ at 4.8 GHz will increase, since 4.8 GHz is 
above $f_w$.

Further reduction in $Q_{\rm rms}$ and $U_{\rm rms}$ 
will result from increasing depolarization
within the source below 8.6 GHz.
As is widely observed, internal Faraday depolarization 
increases strongly with decreasing frequency.
The mean degree of polarization
in PKS 0405-385 decreases below 8.6 GHz,
and so too must the peak level of polarized brightness,
which will reduce the amplitude of polarized ISS.

We interpret the 2.4 and 1.4 GHz IDV as refractive ISS
at frequencies below $f_w$, since the time scales are 
substantially larger than at the two higher frequencies.
In trying to compare these longer time scales with theory, we could
compute the expected time scales for refractive ISS, 
following the R95 method. However, this method does
not include the diffractive ISS contributions at 
frequencies just below $f_w$ and so overestimates the time scale and
underestimates the modulation index for a source as compact as
PKS 0405-385. Simulation techniques are needed to model
the scattering at these frequencies.

In a separate investigation at 1.4 GHz, we considered
the possibility that Faraday depolarization
across the receiver bandpass could have reduced
the peak polarization and so reduced the amplitude
of ISS in $Q$ and $U$. We looked separately
in the 8 sub-bands for $Q,U$ fluctuations
and found no significant differences. 
Thus there is no evidence for
more compact polarized components at 1.4 GHz, which might show more 
rapid (diffractive) ISS in polarization.
This is consistent with the suggestion above that at the lower
frequencies the fine structure in polarized brightness
is suppressed by depolarization in or near the source.

\subsection{Discussion of the Fits}

Whereas two source components are adequate
to model five of the six Stokes' correlations,
a third component is necessary to
adequately model the $QU$ correlation.  Thus
the observed form of $C_{QU,t}(\tau)$ has a strong
influence on the best fitting 3-component source model.
In particular figure \ref{fig:3compx} shows
peaks at lags of -2.0, -0.4 and 0.3 hrs
and accompanying minima at -1.1, -0.2 and 0.8 hrs.
The model shown matches well from -1 to 0.7 hrs, 
but fails to follow the peaks
and valleys further out.  On close inspection we also see
that the negative overshoot in $C_{QQ,t}(\tau)$
and $C_{UU,t}(\tau)$ is also not well modelled.

The model shown has the least rms residual near one local
minimum of the parameter space. There are nearby regions
in which the residual changes little and the peak brightness
increases. However, we did not search exhaustively 
for other local minima far removed from this one.  
In particular we did not explore models with
component separations larger than about 0.04 mas, which
when mapped into time by $\tau = L\theta /V$ corresponds to about 
1.3 hrs.  Thus our model does not include a source sufficiently
extended that could generate peaks at lags of 
2-3 hrs from the origin, such
as those visible in several of the correlations. 
Looking at the correlations over time lags out
to $\pm 7$ hrs in Figure \ref{fig:xc.IQU.cor}, one sees
that the large amplitude correlations do extend out to 
at least 3 hrs.  While we chose to parameterize 
three circular components by their flux density, 
FWHM diameter, relative position, polarized flux density 
and position angle, it is possible that two elliptical
components with gradients in position angle might also give
satisfactory or better fits.

This discussion is relevant also to the lower level of
the polarized ISS at the three lower frequencies
(section \ref{sec:CSL}).
The ISS ``resolution function'' can be
considered to be the function $C_{oo}(\xi,\eta)$ 
(in Figure \ref{fig:ptspread}) mapped into angular coordinates 
by dividing the spatial offsets by the scattering distance $L$. 
This function is elongated having a width equal to the Fresnel angle
($(Lk)^{-0.5}$) in $\theta_x$ 
and a factor four greater in $\theta_y$ for our assumed axial ratio.
If at the lower frequencies there were polarized structures in which
$Q$ or $U$ reversed sign on angles much smaller than
the Fresnel angle, they would tend to average out to give little
observable ISS in $Q$ or $U$ -- that is they would be below the
ISS resolution limit.  We suggest that at the lower frequencies
any polarized structures are subject to greater effects of internal 
Faraday rotation and depolarization, making complex polarization
structure on scales finer than the Fresnel angle and so suppressing
the polarized ISS.  It is interesting that such a resolution limit
does not apply for the ISS of total intensity since it cannot change
sign.  An assembly of sources finer than the Fresnel angle will
still show ISS similar to that of a point source.

\section{Discussion and Conclusions}
\label{sec:disc}

\subsubsection*{Coupling of the Scattering and Source Models}

In Section \ref{sec:dist.dia} we analyzed weak ISS
at 4.8 and 8.6 GHz, caused in a localized layer of 
scattering at distance $L$ from the Earth, for
a single Gaussian source component.
A range  $2 \simless L \simless 30$ pc
matches the observations, with an associated range of 
source diameters $\theta_{\rm mas}$ 
(and a generally inverse relationship between these two parameters).
A major finding of the paper is a reduction in
the peak source brightness temperature to 
$\sim 2 \times 10^{13}$ K, which is a factor 25 less
than inferred in the earlier analysis
of KCJ.  This brings the relativistic bulk Doppler factor 
for the presumed jet closer to the inferred values from
the superluminal motion of VLBI sources.
While we assumed that the IISM velocity was that of the LSR
(36 km s$^{-1}$ relative to the Earth in June),  
a velocity of 75 km s$^{-1}$, 
which is the upper bound from the 
intercontinental time delay in the ISS, 
changes the allowed ranges to 
$8 \simless L \simless 100$ pc, and 
$0.13 \simgreat \theta_{8.6, \rm mas} \simgreat 0.013$ mas
and $2 \times 10^{12} \simless T_b \simless 6 \times 10^{13}$ K,
which is still a factor 8 smaller than inferred by KCJ.

\subsubsection*{Degree of Polarization}

The observations provide values for $Q_{\rm rms}$
and $U_{\rm rms}$, but these were not included in the fitting,
initially, since the normalized correlation functions are unaffected
by the overall degree of polarization. This was
remedied by adding two data points to the computation
of the fitting residual $S^2$.  The squared difference between model 
and observation of $Q_{\rm rms}$ and $U_{\rm rms}$
was added to $S^2$, with a weight such that each would be
approximately equivalent to one of the auto-correlations.
When included in the fit process, this reduced
the peak brightness in the best fitting source models, and
raised the maximum degree of polarization.

In the final two and three component models of Figures
\ref{fig:2comp.polb} and \ref{fig:3comp.polb}
the peak polarization degree is about 70\%,
which is close to the theoretical maximum from a uniform
synchrotron source (see Gardner and Whiteoak 1966).
However, this is not a coincidence, rather it is a result
of an upper bound being placed on the maximum degree
of linear polarization during the fitting process.
The degree of polarization might also
be affected by the addition of a polarized component,
which is too extended to scintillate; depending on its 
position angle this would either increase or decrease the
maximum degree of polarization. However, our model has about
1.3 Jy in a structure that is larger than $\simgreat 0.2$ mas, 
for which the peak brightness is too small to change the 
polarized brightness significantly.
The model shown in Figure \ref{fig:3compx} 
has structure in $Q$ and $U$ on significantly
finer scales than in $I$. Our model is the sum of three
circular Gaussian components, creating a somewhat
elongated structure. The polarized
flux density of each component was constrained to be less than
70\% of the total flux density of component 1. Though
this allows weak Gaussian components with individual degrees
of polarization greater than 100\% (see Table 3), 
it keeps the polarization brightness less than 70 \%
of the total brightness at all points across the source,
consistent with synchrotron theory.
With this constraint the two-component model
gave close agreement in $Q_{\rm rms}$ and $U_{\rm rms}$,
while for the three-component model
these quantities were about 25\% below the observed values.


\subsubsection*{VLBI Observations and Time Evolution of the ISS}

VLB observations of PKS 0405-385 were made in June 1996, revealing 
two components, barely resolved at 1 mas resolution
(Kedziora-Chudczer et al., 2001).  The flux density at 8.4 GHz 
in the ``core'' was 1.5 Jy with 0.24 Jy in a NW extension.
We can ask how such a model might be reconciled 
with the episode of fast IDV in June 1996.
Our results suggest that the IDV was due to 30 by 22 $\mu$as 
compact structure of about 0.5 Jy at 8.6 GHz and
with $\sim 1.3$ Jy in a more extended component (jet?), 
part of which was picked up as a NW extension in the VLB image.
From the observed IDV the compact core has a brightness temperature of
$\sim 2\times 10^{13}$K, which we interpret as 
$\sim 3\times 10^{11}$K Doppler boosted in a relativistic
jet. This implies a minimum Doppler factor of about
14, including the $1+z$ factor ($z = 1.285$)
in Marscher's (1998) expression.
Depending on the angle of the jet to the line of sight 
this corresponds to an apparent superluminal expansion of about 
0.13 mas in two months.  Thus the ultra compact feature causing the ISS
may have expanded from 0.03 mas to 0.13 mas, which could
have quenched the ISS substantially on the time scale of 
two months.  Such behaviour differs considerably from the long-lived 
scintillation seen in the other two very rapid scintillating 
sources, J1819-4835 (DTB, Bignall et al., 2002)

Subsequently the total flux density of the source almost doubled
over 1.5 years indicating further expansion of the previously
scintillating component. In a separate paper we model this long term time
evolution of the PKS 0405-385 as expanding compact knots 
in a more extended jet. However, we cannot yet rule out
the alternative hypothesis that the changes in IDV 
are due to changes in the scattering in the local ISM.

\subsubsection*{The Local ISM}
 
By greatly reducing the scattering distance,
we have greatly reduced the source brightness temperature 
over that inferred by KCJ.
Having replaced a source puzzle by an interstellar puzzle,
we now consider the implications for the
local interstellar medium (ISM)
and look for any corroborating evidence
on the line of sight toward
the quasar ($l=119^{\circ}, b=-48^{\circ}$). 

The $SM$ values in the local ISM 
responsible for the IDV in PKS 0405-385 are 
$\sim 3 \times 10^{-4}$ m$^{-20/3}$ kpc. This 
is comparable to value in the TC93 Galactic disk, 
which extends more than 10 times further than
our screen distance. With a screen thickness of say
5 pc the local value of the scattering strength 
parameter ($C_{N0}^2$) 
must be about 100 times greater
than in the TC93 model. This suggests we
have detected a remarkable nearby concentration 
of very irregular plasma.  

We compare this with
the enhanced scattering at the edge of the 
``local bubble'' (Bhat et al.\ 1998). 
They model their pulsar observations by
an ellipsoidal shell with 
$SM= 6\pm3 \times 10^{-5}$ m$^{-20/3}$ kpc, at 134 pc
in the direction toward PKS 0405-385
(including the factor 3 reduction according to
appendix C of Rickett et al., 2000).
In the set of pulsars that they observed the
closest was 45$^{\circ}$ from our line-of-sight.
Thus taking our observations and theirs together 
requires a much less regular structure than their 
ellipsoidal model of uniform $SM$, 
such that the scattering layer toward PKS 0405-385 has 
a 5 times greater $SM$ and lies at only
25-100 pc from the Earth.  In an approximately orthogonal
direction Rickett et al.\ (2000) deduced a deficit
in scattering on the 310 pc line toward PSR B0809+74
and suggested patchiness in the shell of enhanced
scattering. We can only conclude that the local 
IISM is spatially inhomogeneous
in its turbulence, and we now consider
other evidence on the local ISM toward PKS 0405-385.

The diffuse ionized component of the ISM mapped in the
Southern H$_{\alpha}$ Sky Survey 
(Gaustad et al. 2001) does not reveal any
distinct ionized region in this direction. 
The possibility of a turbulent stellar wind from a
nearby star crossing the line of sight is 
also excluded on the basis of the SUPERCOSMOS
measurements (Miller et al. 1991) and our 
optical imaging with the 1 m telescope at the Mount Stromlo
and Siding Spring Observatory. The images show 
no bright star closer than 1.5', and although they do show
a nearby galaxy at 0.5', the optical spectrum of the quasar
shows an absorption feature at $z = 0.8$ (KCJ), 
but this is too distant to cause the rapid variability.

Various observers have reported observations of
the local interstellar medium
(see Breitschwerdt et al., 1998). 
For example, G\'{e}nova et al.\  (1998) 
report measurements of Na I absorption
lines over a wide range of Galactic longitudes, from which
they deduced the kinematics of several
local interstellar clouds.  They identify an interstellar cloud
``P'' with velocity of 13.8 km~s$^{-1}$ 
(toward $l=225^{\circ}, b=5.4^{\circ}$)
covering a large solid angle that
includes PKS 0405-385.  Since the degree of ionization and
distance to this cloud are not clear, we have no
grounds for identifying it as being associated with the 
enhanced scattering.

Radio observations of the Galactic continuum show the presence
of various arcs and spurs (e.g.\ Haslam et al., 1982). 
Spoelstra (1972) analyzed the radio data in terms of
spherical shells and modelled the ``Cetus Arc''
as a sphere subtending an angular radius of 50$^{\circ}$  
with its center 110 pc from the Earth toward 
$l=110^{\circ}, b=-30^{\circ}$.  The spherical shell
has a radius of 84 pc and thickness $\sim 11$ pc. Though
Spoelstra did not explicitly suggest it this gives
a shortest distance to the inner edge of the 
shell as 26 pc. The direction toward the quasar is 
only $15^{\circ}$ from the center and hence the
distance to the shell is $\sim 27$ pc, which is remarkably
close to the 25 pc used in our model.  This could well be a 
coincidence but nevertheless it points to a possible cause of
the enhanced scattering layer needed to explain the 
rapid ISS observed.
  
\subsubsection*{Polarized IDV at Cm-Wavelength}

Since ISS can explain the very rapid polarized IDV in 0405-385,
one can ask if ISS can explain the generally slower
polarized IDV at centimeter wavelengths from other sources.
Q89 reported rapid polarized IDV from quasar 0917+624, which
was subsequently explained as ISS by R95 and Qian et al.\ (2001).

More recently Gabuzda et al.\ (2000a,b,c) have reported IDV
in polarized flux density and position angle,
observed during VLB observations of three compact sources
at 5 GHz.  They report changes in polarized flux density 
on time scales as short as 4 hours, which are faster and of
larger fractional amplitude than changes in total flux density.
They argue that these changes are intrinsic to particular
components found in the VLB images.  
While intrinsic changes are clearly a viable explanation, 
we suggest the alternative of polarized ISS from
complex very fine structure in the polarized emission
(such as due to a rapid position angle rotation, finer than
the VLBI resolution) and a smoother distribution in 
total brightnes that quenches the ISS in $I$. However,
we have not considered a quantitative ISS model
for their observations.

\subsubsection*{Conclusions}

\begin{itemize}

\item  Weak ISS can explain the rapid IDV at 8.6 GHz and 4.8 GHz,
if it is caused by a local enhancement in scattering
(and turbulence?) at about 25 pc from the Earth.  We have 
assumed that the scattering plasma is stationary in the 
LSR and so we used the velocity of the 
Earth relative to the LSR at the 
time of the observations. At most the velocity might be
twice our assumed value, which would increase the 
scattering distance to 100 pc.

\item The scattering is found to be
highly anisotropic with an axial ratio 1:4, in which the narrow 
dimension of the density micro-structure is within
about $25^{\circ}$ of the effective velocity. 
This provides evidence in
support of strongly anisotropic plasma turbulence as proposed
by Goldreich and Sridhar (1995, 1997), see also 
Nakayama (2001) and Backer and Chandran (2002). 
Further the high degree of anisotropy 
implies a well-ordered magnetic field, 
as might be expected in a relatively thin scattering layer.

\item The peak total brightness temperature in the scintillating component,
which we associate with the compact core of a jet source,
is about $2 \pm 1 \times 10^{13}$ K,   This is comparable to
other highly beamed jet models with Doppler factors 
in the range 14-20.

\item The detailed inter-relations of the linear Stokes' parameters
at 8.6 GHz can be modelled quantitatively by ISS
of a plausible source model in which 0.5 Jy is in an inner
compact component with dimensions $30 \times 22 \mu$as.
The peak degree of polarization is $\sim 70 $\%
and there is a rapid rotation by about $180^{\circ}$ 
of the angle of polarization across the longer axis of the source. 
The results indicate polarized structure on a linear scale of
0.2 pc, which can be compared with parsec scales recently
reported on 12 blazars at 15 and 22 GHz VLBI by Homan et al.\ (2002).
The remaining 1.3 Jy is in a more extended structure larger than,
say, 0.2 mas which may be polarized but is too large to scintillate.
We emphasize that this model is not uniquely determined, but that 
other models with similar features can also be found. 

\item The local enhanced scattering poses a puzzle, which may be resolved
by observations of nearby pulsars. A possible explanation is 
enhanced turbulence thought to exist at the edge
of the local interstellar bubble.  However, in such a case
the boundary of the bubble is quite irregular
and far from a simple ellipsoid.
An alternative is scattering in the nearside 
of an expanding shell, identified
in continuum radio maps as the Cetus arc, which is presumed to be a 
remnant of an expanding supernova shell of about 84 pc in radius.

\item  We have continued regular monitoring of
the the flux density and polarization of the source since 1996,
and will present the results in a future paper,
including the second episode of rapid IDV.
Models for the evolution of the source and of the 
local scattering medium will be discussed.
\end{itemize}

\acknowledgments

The Australia Telescope is funded by the Commonwealth
Government for operation as a National Facility.
The research at UCSD is supported by the US NSF 
under grant AST 9988398. BJR thanks Ron Ekers, the ATNF Director, 
for providing support during his visit to the ATNF, where
this work was completed.

\appendix

\section{Appendix. Correlation Functions: Correction for Noise and Error Analysis}

Errors in the correlation functions result from additive receiver system
(radiometer) noise and from estimation errors due to the 
finite observing span of a stochastic process. 
The receiver noise is effectively ``white'' and independent of the scintillation,
and contributes a spike at zero lag in the auto-correlations. 
As described in section 2, an error was computed for each 70 second
sample of the Stokes parameters, (derived from the rms 
scatter in the 10 second  visibility estimates from the independent 
frequency channels and baselines). Combining these 70 second errors
allows us to estimate the height of the spike, which is then subtracted
from the measured auto-correlation. Thus the auto-correlations displayed 
have been corrected for system noise, though in most cases
the correction is quite small. 

The estimation errors in the auto- and cross-correlations cannot
be removed and are correlated over a range in time lag, 
that is approximately equal to the characteristic time scale
of the data in question. Thus the errors are found in two steps, the
first of which is to estimate the appropriate time scale $\tau$
as defined in section 3. The second step is to calculate the 
number of independent samples of the scintillation in the 
observing time $T_{obs}$. We then approximate the estimation 
error in the normalized auto-correlation at time-lag $t$ by
\be
\sqrt{(1 - \exp[-2 (t/1.2 \tau)^2])  1.25\tau_I /T_{obs} }
\ee
Here the exponential factor is used as an approximation to
the form of the correlation and reflects the fact that the
error in the \it normalized \rm auto-correlation function
goes to zero at zero lag. The noise correction is too small
to invalidate this approximation in most cases.  We used
a similar formula for the error in the cross-correlation functions, 
with the following differences:  the relevant time scale is taken
to be the geometric mean of the two time series involved;
the exponential term is reduced in amplitude by a factor equal to
the cross-correlation coefficient between the two time series.
The point here is that for two time series that are
highly correlated estimates of their mutual cross-correlation
coefficient become increasingly accurate as
their peak correlation approaches unity.  This formulation is
only appropriate for strongly correlated time series
with no systematic time offset.

\section{Weak ISS from Gaussian scattering profile and multiple 
Gaussian source components}
\label{sec:weakscint}

For an extended medium with a Gaussian profile of
scattering strength, we can write the typical cross-spectrum as
\be
P_{IQ}(\kappa_x,\kappa_y) = 8 \pi r_e^2 \lambda^2 
C_{N0}^2    (R \kappa_x^2 + \kappa_y^2/R)^{-\alpha-2} 
Y(\kappa_x,\kappa_y)
\label{eq:PIQY}
\ee
where the $z$ integral is
\be
Y(\kappa_x,\kappa_y) =  \int_0^\infty e^{-(z/L_g)^2)}
V_I(\kappa_x z/2\pi, \kappa_y z/2\pi) V_Q^*(\kappa_x z/2\pi, \kappa_y z/2\pi) 
\sin^2(z \kappa^2 \lambda/4 \pi)  dz  \; .
\label{eq:Y}
\ee

Now consider a source consisting of $nc$ circular Gaussian
components. The visibility in $I$ can be written as
\be
V_I(\vecu ) = \sum_1^{m=nc} S_{I,m} 
\exp[-\pi^2 u^2 \theta_m^2 -2\pi i \vecu .\thetaom ]  \; ,
\ee
where the total flux density in the $m$th component is $S_{I,m}$ with a
radius at ${\rm e}^{-0.5}$ of $\theta_m$ and vector angular
position  $\thetaom$.  Each component
is assumed uniformly linearly polarized, so the visibility
in $Q$ and $U$ are given by the same equations with
$S_{Q,m}$ and $S_{U,m}$ in place of $S_{I,m}$. Here
\be
S_{Q,m} = S_{I,m} p_m \cos(2\chi_m) \;\;\;S_{U,m} = S_{I,m} p_m \sin(2\chi_m)
\ee
Evidently $p_m$ is the fractional polarization and 
$\chi_m$ is the position angle.  In evaluating the extended medium
cross spectrum between $I$ and $Q$ we need
\begin{eqnarray}
V_I V_Q^* = \sum_1^{m=nc} \sum_1^{n=nc} 
S_{I,m} S_{Q,n} \times \nonumber \\
\exp[-\pi^2 u^2 (\theta_m^2-\theta_n^2) -2\pi i \vecu .(\thetaom -\thetaon ) ] \; .
\end{eqnarray}
This can be substituted into equation (\ref{eq:Y}) and the $z$ integral
taken inside the summation so that
\be
Y = \sum_1^{m=nc} \sum_1^{n=nc} Y_{mn} \; ,
\ee
where each term $Y_{mn}$ can be reduced to a sum of integrals
in which the integrand is of the form ${\rm e}^{-a z^2} \cos(bz)$
or ${\rm e}^{-a z^2} \sin(bz)$.  These are given in terms 
exponentials and confluent hypergeometric functions by
standard integrals. The expression for the auto-spectrum 
for a single Gaussian component was given in equation (2)
of Coles et al. (1987)

\end{document}